\documentclass[reprint,amsmath,amssymb,aps]{revtex4-1}
\usepackage{graphicx}
\usepackage{dcolumn}
\usepackage{bm}
\usepackage{color}
\begin{document}

\preprint{APS/123-QED}

\title{ Viscous fluid dynamics with decaying vacuum energy density}

\author{C. P. Singh}
\email[]{cpsingh@dce.ac.in}
\author{Vinita Khatri}
\email[]{vinitakhatri\_2k20phdam501@dtu.ac.in}
\affiliation{Department of Applied Mathematics,\\ Delhi Technological University, Delhi-110042, India}
\date{\today}
\begin{abstract}
In this work, we investigate the dynamics of bulk viscous models with decaying vacuum energy density (VED) in  a spatially homogeneous and isotropic flat Friedmann-Lema\^{i}tre- Robertson-walker (FLRW) spacetime. We particularly are interested to study the viscous models which consider first order deviation from equilibrium, i.e., the Eckart theory. In the first part, using the most general form of bulk viscous coefficient, we find the analytical solutions of main cosmological parameters, like Hubble parameter, scale factor, matter density, deceleration parameter and equation of state parameter and discuss the evolutions of the models accordingly. We also discuss the cosmological consequences of the evolutions and dynamics of three particular viscous models with decaying VED depending on the choices of bulk viscous coefficient. We examine the linear perturbation growth of the model to see if it survives this further level of scrutiny. The second part of the work is devoted to constrain one of the viscous models viz., $\zeta \propto H$, where $\zeta$ is the bulk viscous coefficient and $H$ is the Hubble parameter, using three different combinations of data from type Ia supernovae (Pantheon), $H(z)$ (cosmic chronometers), Baryon Acoustic Oscillation and $f(z)\sigma_8(z)$ measurements with Markov Chain Monte Carlo (MCMC) method. We show that the considered model is compatible with the cosmological probes, and the $\Lambda$CDM recovered in late-time of the evolution of the Universe.  Finally, we obtain selection information criteria (AIC and BIC) to study the stability of the models.
\end{abstract}

\maketitle{}

\section{Introduction}
\label{intro}
 The different observations such as luminosity distances of type Ia supernova, measurements of anisotropy of cosmic microwave background and gravitational lensing  have confirmed that our Universe is spatially flat and expanding with an accelerated rate. It has been observed that the Universe contains a mysterious dominant component, called dark energy (DE) with large negative pressure, which leads to this cosmic acceleration \cite{riess1998,per1999,bennet2003,teg2004,alam2017,amante2019,agh2020}. In literature, several models have been proposed to explain the current accelerated expansion of the Universe. The two most accepted DE models are that of a cosmological constant and a slowly varying rolling scalar field (quintessence models)\cite{cald1998,wang2000,stein2003,peebles2003}.\\
\indent The cosmological constant $\Lambda$(CC for short), initially introduced by Einstein to get the static Universe, is a natural candidate for explaining DE phenomena with equation of state parameter equal to $-1$. The natural interpretation of CC arises as an effect of quantum vacuum energy. Thus, the cold dark matter based cosmology together with a CC, called the $\Lambda$CDM cosmology, is preferred as the standard model for describing the current dynamics of the Universe. It is mostly consistent with the current cosmological observations. However, despite of its success, the $\Lambda$CDM model has several strong problems due to its inability to renormalize the energy density of quantum vacuum, obtaining a discrepancy of $\sim 120$ orders of magnitude between its predicted and observed value, so-called CC or fine-tuning problem \cite{wein1989,carrol2001,padma2003}. It also has the coincidence problem, i.e., why the Universe transition, from decelerated to an accelerated phase, is produced at late times \cite{zlatev1999}.\\
\indent Many models have been proposed to tackle these issues. One of the possible proposal is to incorporate energy transfer among the cosmic components. In this respect, the models with time-varying vacuum energy density (VED), also known as 'decaying vacuum cosmology' seems to be promising. The idea of a time-varying VED models ($\rho_{\Lambda}=\Lambda(t)/8\pi G$) is physically more viable than the constant $\Lambda$ \cite{freese1987,carvalho1992,lima1994,lima1996}. Although no fundamental theory exists to describe a time-varying vacuum, a phenomenological technique has been suggested to parametrize $\Lambda(t)$. In literature, many authors \cite{wang2005,eliza2005,alcaniz2005,borges2005,sola2006,carn2006,borges2008,carn2008,basila2009,basila2009a,costa2010,pigo2011,sola2011,sola2011a,bessada2013,sola2013,per2013,szy2015,sola2015,sola2015a,novikov2016,jaya2019} have carried out analysis on decaying vacuum energy in which the time-varying vacuum has been phenomenologically modeled as a function of time in various possible ways, as a function of the Hubble parameter. Such attempts suggest that decaying VED model provides the possibility of explaining the acceleration of the Universe as well as it solves both cosmological constant and coincidence problems.\\
\indent Shapiro and Sol$\grave{a}$ \cite{sap2002}, and Sol$\grave{a}$ \cite{sola2008} proposed a possible connection between cosmology and quantum field theory on the basis of renormalization group (RG) which gives the idea of running vacuum models (RVM), characterized by VED $\rho_{\Lambda}$, see Refs.\cite{sola2011,sola2013,sola2015a} for a review. The RVM has been introduced to solve the coincidence problem where the term $\Lambda$ is assumed to be varying with the Hubble parameter $H$. Carnerio et al.\cite{carn2008} proposed that the vacuum term is proportional to the Hubble parameter, $\Lambda(a)\propto H(a)$. However, this model fails to fit the current CMB data. It is interesting to note that RG in quantum field theory (QFT) provides a time-varying vacuum, in which $\Lambda(t)$ evolves as $\Lambda \propto H^2$ \cite{grande2006}. Basilakos \cite{basila2009} proposed a parametrization of the functional form of $\Lambda(t)$ by applying a power series expansion in $H$ up to the second order. Recently, a large class of cosmologies has been discussed where VED evolves like a truncated power-series in the Hubble parameter $H$, see Refs.\cite{singh2021,sola2021} and references therein.\\
\indent On the other hand, in recent years, the observations suggest that the Universe is permeated by dissipative fluids. Based on the thermodynamics point of view, phenomenological exotic fluids are supposed to play the role for an alternative DE models. It has been known since long time ago that a dissipative fluid can produce acceleration during the expansion of the Universe \cite{zimdahl2001,balakin2003}. The bulk and shear viscosity are most relevant parts of dissipative fluid. The bulk viscosity characterizes a change in volume of the fluid which is relevant only for the compressed fluids. The shear viscosity characterizes a change in shape of a fixed volume of the fluid which represents the ability of particles to transport momentum. In general, shear viscosity is usually used in connection with the spacetime anisotropy where as bulk viscosity plays the role in an isotropic cosmological models. The dynamics of homogeneous cosmological models has been studied in the presence of viscous fluid and has application in studying the evolution of the Universe. \\
 \indent Eckart \cite{eckart1940} extended a classical irreversible thermodynamics from Newtonian to relativistic fluids. He proposed the simplest non-causal theory of relativistic dissipative phenomena of first order which was later modified by Landau and Lifshitz \cite{landau1987}. The Eckart theory has some important limitations. It has been found that all the equilibrium states are unstable \cite{hiscock1985} and the signals can propagate through the fluids faster than the speed of light \cite{muller1967}. Therefore, to resolve theses issues, Israel and Stewart \cite{stewart1976} proposed a full causal theory of second order. When the relaxation time goes to zero, the causal theory reduces to the Eckart's first order theory. Thus, taking the advantage of this limit of vanishing relaxation time at late time, it has been used widely to describe the recent accelerated expansion of the Universe. An exhaustive reviews on non-causal and causal theories of viscous fluids can be found in Refs.\cite{gron1990,maartens1996,coley1996,zimdahl1996,brevik2002,brevik2004,singh2007,singh2008,nojori2007,brevik2010,velten2013,
wang2014,bamba2016,brevik2017}. In recent years, the direct observations indicate for viscosity dominated late epoch of accelerating expansion of the Universe. In this respect, many authors have explored the viability of a bulk viscous Universe to explain the present accelerated expansion of the Universe cf.\cite{kremer2003,fabris2006,brevik2005,hu2006,ren2006,meng2007,wilson2007,mathews2008,avelino2008,avelino2008a,avelino2009,meng2009,avelino2010,
singh2014,mathew2015,normann2017,wang2018,singh2018,singh2018a,singh2018b,singh2019,singh2020}.\\
\indent In Eckart theory, the effective pressure of the cosmic fluid is modeled as $\Pi=-3\zeta H$, where $\zeta$ is bulk viscous coefficient and $H$ the Hubble parameter. Bulk viscous coefficient can be assumed as a constant or function of Hubble parameter. It allows to explore the presence of interacting terms in the viscous fluid. Since the imperfect fluid should satisfy the equilibrium condition of thermodynamics, the pressure of the fluid must be greater than the one produced by the viscous term. To resolve this condition, it is useful to add an extra fluid such as cosmological constant. Many authors \cite{singh2007a,singh2009,nour2011,hu2020,herr2020} have studied viscous cosmological models with constant or with time-dependent cosmological constant. Hu and Hu \cite{hu2020} have investigated a bulk viscous model with cosmological constant by assuming bulk viscous proportional to the Hubble parameter. Herrera-Zamorano et al. \cite{herr2020} have studied a cosmological model filled with two fluids under Eckart formalism, a perfect fluid as DE mimicking the dynamics of the CC, while a non-perfect fluid as dark matter with viscosity term. \\
\indent In this paper, we focus on discussing the dynamics of viscous Universe which consider the first order deviation from equilibrium, i.e., Eckart formalism with decaying VED. Using different versions of bulk viscous coefficient $\zeta$, we find analytically the main cosmological functions such as the scale factor, Hubble parameter, matter  density, deceleration and equation of state parameters. We discuss the effect of viscous model with varying VED in perturbation level. We implement the perturbation equation to obtain the growth of matter fluctuations in order to study the contribution of this model in structure formation. We perform a Bayesian Markov Chain Monte Carlo (MCMC) analysis to constrain the parameter spaces of the model using three different combinations involving observational data from type Ia supernovae (Pantheon), Hubble data (cosmic chronometers), Baryon acoustic oscillations and $f(z)\sigma_8(z)$ measurements . We compare our model and concordance $\Lambda$CDM to understand the effects of viscosity with decaying vacuum by plotting the evolutions of the deceleration parameter, equation of state parameter and Hubble parameter. We also study the selection information criterion such as AIC and BIC to analyze the stability of the model. \\
\indent The work of the paper is organized as follows. In Section II, we present the basic cosmological equations of Friedmann-Lema\^{i}tre-Robertson-Walker (FLRW) geometry with bulk viscosity and decaying VED. In Section III, we find the solution of the field equations by assuming the most general form of bulk viscous coefficient. Section IV is devoted to study the evolutions of some particular forms of bulk viscous coefficient with varying VED. We discuss the growth rate equations that govern the perturbation in Section V. Section VI presents the observational data and method to be used to constrain the proposed model. The results and discussion on the evolution of the various parameters are presented in Section VII. In Section VIII, we present the selection information criterion to distinguish the presented model with concordance $\Lambda$CDM. Finally, we conclude our finding in Section IX.
\section{Viscous model with varying-$\Lambda$}
\label{sec:1}
Let us start with the Friedmann-Lemaitre-Robertson-Walker (FLRW) metric in the flat space geometry as the case favoured by observational data
\begin{equation}\label{eq1}
ds^2 = -dt^2 + a^2(t) \left[dr^2+r^2(d\theta^2+sin^2\theta d\phi^2)\right],
\end{equation}
where $(r,\theta,\phi)$ are the co-moving coordinates and $a(t)$ is the scale factor of the Universe. The large scale dynamics of \eqref{eq1} is described by the Einstein field equations, which include the cosmological constant $\Lambda$ and is given by
\begin{equation}\label{eq2}
G_{\mu\nu}=R_{\mu\nu}-\frac{1}{2}g_{\mu\nu}R =8\pi G(T_{\mu\nu}+g_{\mu\nu}\rho_{\Lambda}),
\end{equation}
where $G_{\mu\nu}$ is the Einstein tensor, $\rho_{\Lambda}=\Lambda/8\pi G$ is the vacuum energy density (the energy density associated to CC vacuum term) and $T_{\mu\nu}$ is the energy-momentum tensor of matter. It is to be noted that for simplicity we use geometrical units $8\pi G=c=1$. We introduce a bulk viscous fluid through the energy-momentum tensor which is given by \cite{weinberg1971}
\begin{equation}\label{eq3}
T_{\mu\nu}=(\rho_{m}+P)u_{\mu}u_{\nu}+g_{\mu\nu}P,
\end{equation}
where $u^{\mu}$ is the fluid four-velocity, $\rho_m$ is the density of matter and $P$ is the pressure which is composed of the barotropic pressure $p_m$ of the matter fluid plus the viscous pressure $\Pi$, i.e., $P=p_{m}+\Pi$. The origin of bulk viscosity is assumed as a deviation of any system from the local thermodynamic equilibrium. According to the second law of thermodynamics, the re-establishment to the thermal equilibrium is a dissipative processes which generates entropy. Due to generation of entropy, there is an expansion in the system through a bulk viscous term.\\
\indent In homogeneous and isotropic cosmological models, the viscous fluid is characterized by a bulk viscosity. It is mostly based on the Eckart's formalism \cite{eckart1940} which can be obtained from the second order theory of non-equilibrium thermodynamics proposed by Israel and Stewart \cite{stewart1976} in the limit of vanishing relaxation time.  The viscous effect can be defined by the viscous pressure $\Pi=-3\zeta H$, where $\zeta$ is the bulk viscous coefficient and $H$ is the Hubble parameter. The bulk viscous coefficient $\zeta$ is assumed to be positive on thermodynamical grounds. Therefore, it makes the effective pressure as a negative value which leads to modification in energy-momentum tensor of perfect fluid.\\
\indent If we denote the total energy-momentum tensor $T_{\mu\nu}+g_{\mu\nu}\rho_{\Lambda}$ as modified $\tilde{T}_{\mu\nu}$ on right hand side of field equations \eqref{eq2}, then the modified $\tilde{T}_{\mu\nu}$ can be assumed the same form as $T_{\mu\nu}$, that is, $\tilde{T}_{\mu\nu}=(\rho+p)u_{\mu}u_{\nu}+g_{\mu\nu}p$, where $\rho=\rho_{m}+\rho_{\Lambda}$ and $p=p_m-3\zeta H+p_{\Lambda}$ are the total energy density and pressure, respectively. Further, we assume that the bulk viscous fluid is the non-relativistic matter with $p_m=0$. Thus, the contribution to the total pressure is only due to the sum of negative viscous pressure, $-3\zeta H$ and vacuum energy pressure, $p_{\Lambda}=-\rho_{\Lambda}$. \\
 \indent Using the modified energy-momentum tensor as discussed above, the Einstein field equations \eqref{eq2} describing the evolution of FLRW Universe dominated by bulk viscous matter and vacuum energy yield
\begin{equation}\label{eq4}
3H^2= \rho= \rho_m+\rho_{\Lambda},
\end{equation}
\begin{equation}\label{eq5}
2\dot{H}+3H^2=-p=3\zeta H + \rho_{\Lambda}.
\end{equation}
\noindent where $H=\dot{a}/a$ is the Hubble parameter and an over dot represents the derivative with respect to cosmic time $t$. \indent In this paper, we propose the evolution of the Universe based on decaying vacuum models, i.e., vacuum energy density as a function of the cosmic time. From \eqref{eq2}, the Bianchi identity $\nabla^{\mu}G_{\mu\nu}=0$ gives
\begin{equation}\label{eq6}
\nabla^{\mu}\tilde{T}_{\mu\nu}=0,
\end{equation}
or, equivalently,
\begin{equation}\label{eq7}
\dot{\rho}_{m}+3H(\rho_{m}+p_{m}-3\zeta H+\rho_{\Lambda}+p_{\Lambda})=-\dot{\rho}_{\Lambda},
\end{equation}
which imply that the there is a coupling between a dynamical $\Lambda$ term and viscous CDM. Therefore, there is some energy exchange between the viscous CDM fluid and vacuum. Using the equation of state of the vacuum energy  $p_{\Lambda}=-\rho_{\Lambda}$ and $p_{m}=0$, Eq. \eqref{eq7} leads to
\begin{equation}\label{eq8}
\dot{\rho}_{m}+3H(\rho_{m}-3\zeta H)=-\dot{\rho}_{\Lambda}.
\end{equation}
Now, combining \eqref{eq4} and \eqref{eq8}, we get
\begin{equation}\label{eq9}
\dot{H}+\frac{3}{2} H^2=\frac{1}{2}\rho_{\Lambda}+\frac{3}{2}\zeta H.
\end{equation}
The evolution equation \eqref{eq9} has three independent unknown quantities, namely, $H$, $\zeta$ and $\rho_{\Lambda}$. We get the solution only if $\zeta$ and $\rho_{\Lambda}$ are specified. In what follows, we discuss the dynamics of the Universe depending on the specific forms of $\rho_{\Lambda}$ and $\zeta$.
\section{Solution of field equations}
In this paper, we parameterize the functional form of $\rho_{\Lambda}$ as a function of Hubble parameter. The motivation for a function $\rho_{\Lambda}=\rho_{\Lambda}(H)$ can be assumed from different points of view. Although the correct functional form of $\rho_{\Lambda}$ is not known, a quantum field theory (QFT) approach within the context of the renormalization group (RG) was proposed in Refs.\cite{adler1982,parker1985} and further studied by many authors \cite{basila2009a,sola2011,sola2013,sola2008,shapiro2009,costa2012}. In Ref. \cite{per2013}, the following ratio has been defined between the two fluid components:
\begin{equation}\label{eq10}
\gamma=\frac{\rho_{\Lambda}-\rho_{\Lambda_0}}{\rho_m+\rho_{\Lambda}},
\end{equation}
where $\rho_{\Lambda_0}$ is a constant vacuum density. If $\rho_{\Lambda}=\rho_{\Lambda_0}$, then $\gamma=0$, and we get $\Lambda$CDM model. On the other hand, if $\rho_{\Lambda_0}\neq0$, then we get
\begin{equation}\label{eq11}
\rho_{\Lambda}=\rho_{\Lambda0}+\gamma (\rho_m+\rho_{\Lambda})=\rho_{\Lambda0}+3\gamma H^2.
\end{equation}
The above proposal was first considered by Shapiro and Sola \cite{sap2002} in context of RG. Many authors have studied the evolution of the Universe by assuming this form \cite{sola2011a,bessada2013,jaya2019}. Hereafter, we shall focus on the simplest form of $\rho_{\Lambda}$ which evolves with the Hubble rate. Specifically, in this paper we consider
\begin{equation}\label{lam}
\rho_{\Lambda}=c_0+3\nu H^2,
\end{equation}
where $c_0=3H^2_0(\Omega_{\Lambda 0}-\nu)$ is fixed by the boundary condition $\rho_{\Lambda}(H_0)=\rho_{\Lambda 0}$. The suffix `$0$' denotes the present value of the parameter. The dimensionless coefficient $\nu$ is the vacuum parameter and  is expected to be very small value $|\nu|\ll1$. A non-zero value of it makes possible the cosmic evolution of the vacuum.\\
\indent The choice of $\zeta$ generates different viscous models and in literature there are different approaches to assume the evolution of bulk viscosity. In this paper, we consider the most general form of the bulk viscous term $\zeta$ , which is assumed to be the sum of three terms: the first term is a constant, $\zeta_0$, the second term is proportional to the Hubble parameter $H=\dot{a}/{a}$ which is related to the expansion and the third term is proportional to the acceleration, $\ddot{a}/\dot{a}$. Thus, we assume the parametrization of bulk viscous coefficient in the form\cite{brevik2017,ren2006,meng2007,meng2009,mathew2015,avelino2013,sasidharan2016}
\begin{equation}\label{eq12}
\zeta=\zeta_0+\zeta_1\frac{\dot{a}}{a}+\zeta_2 \frac{\ddot{a}}{\dot{a}},
\end{equation}
where $\zeta_0$, $\zeta_1$ and $\zeta_2$ are constants to be determined by the observations. The term $\ddot{a}/\dot{a}$ in Eq. \eqref{eq12} can be written as $\ddot{a}/aH$. The basic idea about the assumption of $\zeta$ in Eq.\eqref{eq12} is that the dynamic state of the fluid influences its viscosity in which the transport viscosity is related to the velocity and acceleration.  \\
\indent  Using Eqs.\eqref{lam} and \eqref{eq12}, the differential equation for the Hubble parameter \eqref{eq9} finally reduces to
\begin{equation}\label{eqd1}
\left(1-\frac{3}{2}\zeta_2\right)\dot{H}+\frac{3}{2}\left(1-\zeta_1-\zeta_2-\nu\right)H^2-\frac{3}{2}\zeta_0H-\frac{1}{2}c_0=0,
\end{equation}
which on integration, it gives
\begin{equation}\label{x1}
H=\frac{\zeta_0}{2(1-\zeta_1-\zeta_2-\nu)}+\sigma \left(\frac{1+e^{-3(1-\zeta_1-\zeta_2-\nu)\sigma t}}{1- e^{-3(1-\zeta_1-\zeta_2-\nu)\sigma t}}\right),
\end{equation}
where $\sigma=\sqrt{\left(\frac{\zeta_0}{2(1-\zeta_1-\zeta_2-\nu)}\right)^2+\frac{H^2_0 ( \Omega_{\Lambda0}-\nu)}{(1-\zeta_1-\zeta_2-\nu)}}$.\\
\indent The above equation simplifies to give
\begin{equation}\label{eqd2}
H=\frac{\zeta_0}{2(1-\zeta_1-\zeta_2-\nu)}+\sigma \coth \left( \frac{3}{2}\frac{(1-\zeta_1-\zeta_2-\nu)\sigma}{(1-\frac{3}{2}\zeta_2)}t\right).
\end{equation}
Using the Hubble parameter $H=\dot{a}/a$, the scale factor of the model $a(t)$ with the condition $a(t_0)=1$ is given by
\begin{equation}\label{eqd3}
a=e^{\frac{\zeta_0}{2(1-\zeta_1-\zeta_2-\nu)}t}\left[\sinh\left( \frac{3}{2}\frac{(1-\zeta_1-\zeta_2-\nu)\sigma}{(1-\frac{3}{2}\zeta_2)}t\right)\right]^{\frac{2(1-\frac{3}{2}\zeta_2)}{3(1-\zeta_1-\zeta_2-\nu)}},
\end{equation}
which shows that the scale factor increases exponentially as $t$ increases. From \eqref{eqd3}, one can observe that, in general, it is not possible to express cosmic time $t$ in terms of the scale factor $a$. It is possible only if the viscous coefficient terms are zero. In the absence of bulk viscosity, we obtain the result of decaying vacuum model as discussed in Ref.\cite{basila2009a}. Further, for constant $\Lambda$, the solution reduced to the $\Lambda$CDM model with no viscosity. \\
\indent It is  worthwhile to compute the evolution of matter energy density as a function of scale factor (or redshift) or function of cosmic time. Using \eqref{lam} and \eqref{eq12}, the continuity equation \eqref{eq8} takes the form
\begin{equation}\label{ce1}
\dot{\rho_m}+3(1-\nu)H \rho_m=9(1-\nu)\left(\zeta_0+\zeta_1\frac{\dot{a}}{a}+\zeta_2 \frac{\ddot{a}}{\dot{a}}\right)H^2.
\end{equation}
The solution of above equation involves a big expression. Therefore, we avoid to write the expression for matter density. However, we present the numerical solution of this equation for different combinations of viscous and $\nu$ terms in Fig.1. It is observed that the matter energy density decreases as $t$ increases and it approaches to the finite value as $t\rightarrow \infty$. However, in the absence of viscous terms, $\rho_m \rightarrow 0$ as $t\rightarrow \infty$. \\
\begin{figure}[t]
\includegraphics[width=8.0cm, height=6.0cm]{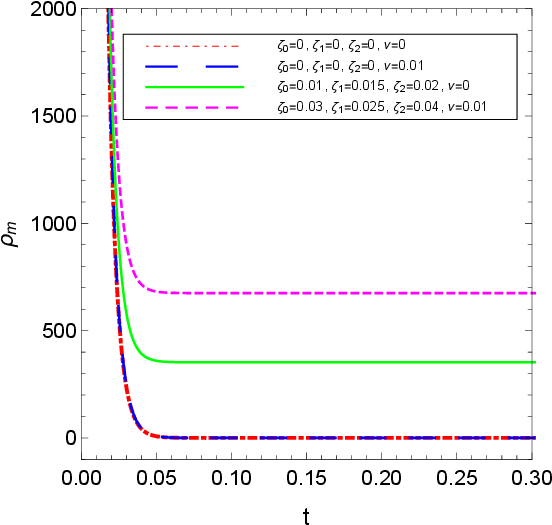}
\caption{The matter energy density as a function of cosmic time $t$ for decaying vacuum with $\zeta=\zeta_0+\zeta_1\frac{\dot{a}}{a}+\zeta_2 \frac{\ddot{a}}{\dot{a}}$}.
\label{fig:1}
\end{figure}
\indent To discuss the decelerated and accelerated phases, and its transition during the evolution of the Universe, we study a cosmological parameter, known as `deceleration parameter', $q$, which is defined as
\begin{equation}\label{x2}
q=-\frac{\ddot{a}}{a}\frac{1}{H^2}=-\left(1+\frac{\dot{H}}{ H^2}\right).
\end{equation}
Using \eqref{eqd2}, the deceleration parameter is calculated as
\begin{equation}\label{x3}
q=-1
 +\frac{3}{2}\frac{\frac{(1-\zeta_1-\zeta_2-\nu)}{(1-\frac{3}{2}\zeta_2)}\sigma^2 \csc^2h (\frac{3}{2}\frac{(1-\zeta_1-\zeta_2-\nu)}{(1-\frac{3}{2}\zeta_2)}\sigma t)}{\left(\frac{\zeta_0}{2(1-\zeta_1-\zeta_2-\nu)}+\sigma \coth(\frac{3}{2}\frac{(1-\zeta_1-\zeta_2-\nu)}{(1-\frac{3}{2}\zeta_2)}\sigma t)\right)^2}.
\end{equation}
From \eqref{x3}, we observe that the model transits from decelerated phase to accelerated phase. As $t$ increases, the deceleration parameter decreases and as $t\rightarrow \infty$, it approaches to $q=-1$. The rate of deceleration parameter attaining to $-1$ depends on the viscous terms.\\
\indent For sake of completeness, we discuss another important cosmological parameter, known as effective equation of state (EoS) parameter, which is defined as
\begin{equation}\label{eq19}
w_{eff}=-1-\frac{2}{3}\frac{\dot{H}}{H^2}.
\end{equation}
Using \eqref{eqd2}, Eq.\eqref{eq19} gives
\begin{equation}\label{eqc6}
w_{eff}=-1
 +\frac{\frac{(1-\zeta_1-\zeta_2-\nu)}{(1-\frac{3}{2}\zeta_2)}\sigma^2 \csc^2h (\frac{3}{2}\frac{(1-\zeta_1-\zeta_2-\nu)}{(1-\frac{3}{2}\zeta_2)}\sigma t)}{\left(\frac{\zeta_0}{2(1-\zeta_1-\zeta_2-\nu)}+\sigma \coth(\frac{3}{2}\frac{(1-\zeta_1-\zeta_2-\nu)}{(1-\frac{3}{2}\zeta_2)}\sigma t)\right)^2}.
\end{equation}
It can be observed that the effective EoS parameter decreases to negative values and finally saturated to $w_{eff}=-1$ corresponding to a de Sitter epoch in future time of evolution.\\
\section{Some particular solutions}
In order to calculate specific expressions for cosmological parameters of viscous model with decaying vacuum energy, let us analyze three particular popular proposals depending on the choice of $\zeta$ defined in Eq.\eqref{eq12} \\

\noindent \textbf{\it Case I:  $\zeta=\zeta_0$=const.}\\

This is the simplest parametrization of Eckart's bulk viscosity model. Many authors \cite{brevik2005,hu2006,avelino2009,singh2018,singh2019,singh2020,nour2011,ajay2019,chitre1987,montiel2011} have studied the viscous cosmological models with constant bulk viscous coefficient. In this case, the evolution equation \eqref{eqd1} reduces to
\begin{equation}\label{eq13}
\dot{H}+\frac{3}{2}(1-\nu)H^2-\frac{3}{2}\zeta_0 H=\frac{1}{2}c_0.
\end{equation}
Solving \eqref{eq13} or directly taking $\zeta_1=\zeta_2=0$ in \eqref{eqd2}, for $\nu <1$, we get
\begin{equation}\label{eq15}
H=\frac{\zeta_0}{2(1-\nu)}+\sigma_1 \coth\left(\frac{3}{2}(1-\nu)\sigma_1 t\right),
\end{equation}
where \\\\$\sigma_1=\sqrt{(\frac{\zeta_0}{2(1-\nu)})^2+\frac{H^2_0(\Omega_{\Lambda0}-\nu)}{(1-\nu)}}$.  It can be observed that the solution reduces to the standard $\Lambda$ for $\zeta_0=0$ and $\nu=0$, whereas for $\zeta_0=0$ and $\nu \ne 0$ it gives the solution for $\Lambda(t)$ model from quantum field theory\cite{basila2009a}. The scale factor is given by
\begin{equation}\label{eq16}
a(t)=e^{\frac{\zeta_0}{2(1-\nu)}t}\left(\sinh(\frac{3}{2}(1-\nu)\sigma_1 t)\right)^{\frac{2}{3(1-\nu)}},
\end{equation}
which shows that the scale factor increases exponentially as $t$ increases. From \eqref{eq16}, one can observe that, in general, it is not possible to express cosmic time $t$ in terms of the scale factor $a$. It is possible only if $\zeta_0=0$. In the absence of bulk viscosity, we obtain the result of decaying vacuum model as discussed in Ref.\cite{basila2009a}. Further, for constant $\Lambda$, the solution reduced to the $\Lambda$CDM model with no viscosity. \\
\indent The deceleration parameter and effective EoS parameter are respectively given by
\begin{equation}\label{eq18}
q=-1+\frac{3}{2}\frac{(1-\nu)\sigma_1^2 \csc^2h (\frac{3}{2}(1-\nu)\sigma_1 t)}{\left(\frac{\zeta_0}{2(1-\nu)}+\sigma_1 \coth(\frac{3}{2}(1-\nu)\sigma_1 t)\right)^2},
\end{equation}
\noindent and\\
\begin{equation}\label{eq20}
w_{eff}=-1+\frac{(1-\nu)\sigma_1^2 \csc^2h (\frac{3}{2}(1-\nu)\sigma_1 t)}{\left(\frac{\zeta_0}{2(1-\nu)}+\sigma_1 \coth(\frac{3}{2}(1-\nu)\sigma_1 t)\right)^2}.
\end{equation}
The time evolutions of $q$ and $w_{eff}$ are similar to the evolutions of these parameters as discussed  for general form of viscous term in Sect.III. \\
\indent The continuity equation \eqref{ce1} in this case has the form
 \begin{equation}\label{case1}
 \dot{\rho_m}+3(1-\nu)H\rho_m=9(1-\nu)\zeta_0H^2
 \end{equation}
Solving \eqref{case1}, one may find the time evolution of the matter density.  We will only present a numerical solution of this equation. In Fig.2 we plot the time evolution of matter energy density $\rho_m(t)$ for different combinations of $\zeta_0$ and $\nu$. This figure shows that the matter density diverges at beginning of the cosmic evolution and decreases as $t$ increases, and finally approaches to a finite value as $t\rightarrow \infty$ for $\zeta \ne 0$ and $\nu \ne 0$. In the absence of viscosity or decaying vacuum energy, the matter energy density tends to zero as $t\rightarrow \infty$.\\

\noindent \textbf{\it Case II:  $\zeta=\zeta_0+\zeta_1 H$}\\

We assume that the bulk viscous coefficient is a linear combination of two terms: $\zeta_0$ and $\zeta_1 H$, i.e., $\zeta=\zeta_0+\zeta_1H$. In literature, many authors \cite{meng2007,meng2009,avelino2010} have assumed such a form of $\zeta$ to study the dynamics of Universe.  In this case, Eq.\eqref{eqd1} reduces to
\begin{equation}\label{eqc1}
\dot{H}+\frac{3}{2} (1-\zeta_1-\nu)H^2-\frac{3}{2}\zeta_0 H=\frac{1}{2}c_0.
\end{equation}
Solving Eq. \eqref{eqc1} or directly putting $\zeta_2=0$ in Eq.\eqref{eqd2},  the solution for Hubble parameter for $(\zeta_1+\nu)<1$ is given by
\begin{figure*}
\begin{minipage}[t]{0.40\linewidth}
\includegraphics[width=\linewidth]{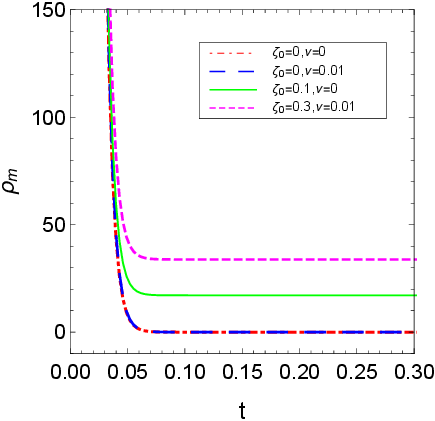}
\caption{The time evolution of matter energy density for decaying vacuum model with viscosity $\zeta=\zeta_0$.}\label{fig:1}
\end{minipage}
\hfill
\begin{minipage}[t]{0.40\linewidth}
\includegraphics[width=\linewidth]{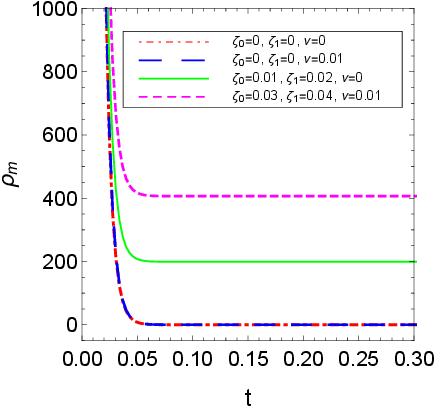}
\caption{The time evolution of matter energy density for decaying vacuum model with viscosity $\zeta=\zeta_0+\zeta_1 H$ }\label{fig:2}
\end{minipage}
\end{figure*}
\begin{equation}\label{eqc3}
H=\frac{\zeta_0}{2(1-\zeta_1-\nu)}+\sigma_2\; \coth\left(\frac{3}{2}(1-\zeta_1-\nu)\sigma_2 t\right),
\end{equation}
where $\sigma_2=\sqrt{\left(\frac{\zeta_0}{2(1-\zeta_1-\nu)}\right)^2+\frac{H^2_0 (\Omega_{\Lambda0}-\nu)}{(1-\zeta_1-\nu)}}$. The corresponding expression for the scale factor in normalized unit has the form
\begin{equation}\label{eqc4}
a=e^{\frac{\zeta_0}{2(1-\zeta_1-\nu)}t}\left[\sinh\left(\frac{3}{2}(1-\zeta_1-\nu)\sigma_2 t\right)\right]^{\frac{2}{3(1-\zeta_1-\nu)}}.
\end{equation}
The respective deceleration parameter and effective EoS parameter are calculated as
\begin{equation}\label{eqc5}
q=-1
 +\frac{3(1-\zeta_1-\nu)\sigma_2^2 \csc^2h (\frac{3}{2}(1-\zeta_1-\nu)\sigma_2 t)}{2\left(\frac{\zeta_0}{2(1-\zeta_1-\nu)}+\sigma_2 \coth(\frac{3}{2}(1-\zeta_1-\nu)\sigma_2 t)\right)^2},
\end{equation}
and\\
\begin{equation}\label{eqc6}
w_{eff}=-1+\frac{(1-\zeta_1-\nu)\sigma_2^2 \csc^2h (\frac{3}{2}(1-\zeta_1-\nu)\sigma_2 t)}{\left(\frac{\zeta_0}{2(1-\zeta_1-\nu)}+\sigma_2 \coth(\frac{3}{2}(1-\zeta_1-\nu)\sigma_2 t)\right)^2}.
\end{equation}
The time evolutions of $q$ and $w_{eff}$ are similar to the evolutions of these parameters as discussed  for general form of viscous term in Sect.III. \\
\indent The continuity equation \eqref{ce1} in this case has the form
 \begin{equation}\label{case2}
 \dot{\rho_m}+3(1-\nu)H\rho_m=9(1-\nu)(\zeta_0H^2+\zeta_1H^3).
 \end{equation}
We only present a numerical solution of Eq.\eqref{case2}. In Fig.3 we plot the time dependent matter energy density $\rho_m(t)$ for different combinations of $\zeta_0$, $\zeta_1$ and $\nu$. It is observed from figure that the matter density diverges at beginning of the cosmic evolution and decreases as time passes, and finally approaches to a finite value as $t\rightarrow \infty$ for $\zeta \ne 0$,  $\zeta_1 \ne 0$ and $\nu \ne 0$. In the absence of viscous terms or decaying vacuum energy, the matter energy density tends to zero as $t\rightarrow \infty$.\\

\noindent \textbf{\it Case III : $\zeta=\zeta_1 H$}\\

Finally, let us consider the case where bulk viscous coefficient is proportional to the Hubble parameter, i.e., $\zeta=\zeta_1 H$. Such a form of $\zeta$ has been studied by many authors \cite{gron1990,meng2007,singh2007,singh2014,brevik2002a,huan2020}. In this case, the evolution equation \eqref{eqd1} for Hubble parameter reduces to
\begin{equation}\label{eqb1}
\dot{H}+\frac{3}{2}(1-\zeta_1-\nu)H^2-\frac{1}{2}c_0=0.
\end{equation}
The above equation with change of a variable from $t$ to $x=\ln a$ can be written as
\begin{equation}\label{eqb2}
\frac{dh^2}{dx}+3(1-\zeta_1-\nu)h^2=3(\Omega_{\Lambda0}-\nu),
\end{equation}
where $h=H/H_0$ is the dimensionless Hubble parameter and $\Omega_{\Lambda0}=\rho_{\Lambda0}/3H^2_0$. Assuming $(\zeta_1+\nu)<1$ and using the normalized scale factor -redshift relation, $a=(1+z)^{-1}$,  we can express the normalized Hubble function $E(z)\equiv H(z)/H_0$ as
\begin{align}\label{eqb3}
E(z)&=\frac{1}{(1-\zeta_1-\nu)^{1/2}}\nonumber\\
& \times\left[(1-\zeta_1-\Omega_{\Lambda 0})(1+z)^{3(1-\zeta_1-\nu)}+ \Omega_{\Lambda 0}-\nu \right]^{1/2}.
\end{align}
From the above equation, it is clear that for $\nu=0$ and $\zeta_1=0$, we recover exactly the $\Lambda$CDM expansion model whereas only $\zeta_1=0$ gives the solution obtained in Ref.\cite{jaya2019}. It is observed that at very late time we get an cosmological constant dominated era, $H\approx H_0\sqrt{\frac{\Omega_{\Lambda0}-\nu}{(1-\zeta_1-\nu)}}$, which implies a de Sitter phase of the scale factor. Using $H=\dot{a}/a$, the solution for the scale factor in terms of cosmic time $t$ is given by
\begin{align}\label{eqb4}
a&=\left(\frac{(1-\zeta_1-\Omega_{\Lambda0})}{\Omega_{\Lambda0}-\nu}\right)^{\frac{1}{3(1-\zeta_1-\nu)}}\nonumber \\
& \times\left[\sinh(\frac{3}{2}\sqrt{(1-\zeta_1-\nu)(\Omega_{\Lambda0}-\nu)}\;H_0\;t)\right]^{\frac{2}{3(1-\zeta_1-\nu)}}
\end{align}
It can be observed that the scale factor evolves as power-law expansion, i.e., $a\propto t^{2/3(1-\zeta_1-\nu)}$ for small values of $t$ whereas it expands exponentially, i.e., $a\propto \exp{\sqrt{\frac{(\Omega_{\Lambda0}-\nu)}{3(1-\zeta_1-\nu)}}H_0 t}$ for large values of time $t$. In other words, the model expands with decelerated rate in early time of its evolution and expands with accelerated rate in late time of its evolution. \\
\indent From Eq. \eqref{eqb4}, we can find the cosmic time in terms of the scale factor, which is given by
\begin{equation}\label{eqb5}
t(a)=\frac{2}{3H_0\sqrt{(1-\zeta_1-\nu)(\Omega_{\Lambda0}-\nu)}}\sinh^{-1}\left[\left(\frac{a}{a_I}\right)^{\frac{3(1-\zeta_1-\nu)}{2}}\right]
\end{equation}
where $a_I=\left(\frac{(1-\zeta_1-\Omega_{\Lambda0})}{(\Omega_{\Lambda0}-\nu)}\right)^{1/3(1-\zeta_1-\nu)}$.\\
\indent Using \eqref{eqb3}, the value of $q$ in terms of redshift is calculated as
\begin{equation}\label{eqb7}
q(z)=-1+\frac{3}{2}\frac{(1-\zeta_1-\Omega_{\Lambda0})(1+z)^{3(1-\zeta_1-\nu)}}{\left[\frac{(\Omega_{\Lambda0}-\nu)}{(1-\zeta_1-\nu)}+\left(1-\frac{(\Omega_{\Lambda0}-\nu)}{(1-\zeta_1-\nu)}\right)(1+z)^{3(1-\zeta_1-\nu)}\right]}
\end{equation}
The above equation shows that the dynamics of $q$ depends on the redshift which describes the transition of the Universe from decelerated to accelerated phase. We observe that as $z\rightarrow -1$, $q(z)$ approaches to $-1$. However, the model decelerates or accelerates if $\Omega_{\Lambda0}=\nu$, which gives $q=-1+1.5(1-\zeta_1-\nu)$. Thus, a cosmological constant is required for a transition phase. Also, for $z=0$, we find the present value of $q$ which is given by
\begin{equation}\label{eqb8}
q_0=-1+1.5(1-\zeta_1-\Omega_{\Lambda0}).
\end{equation}
\indent  The transition redshift, $z_{tr}$ of the Universe, which is defined as a zero point of the deceleration parameter, $q=0$, can be calculated as
\begin{equation}\label{eqb9}
z_{tr}=-1+\left(\frac{2(\Omega_{\Lambda0}-\nu)}{(3(1-\zeta_1-\nu)-2)(1-\zeta_1-\Omega_{\Lambda0})}\right)^{\frac{1}{3(1-\zeta_1-\nu)}}.
\end{equation}
\indent In this case, the effective EoS parameter is defined by $w_{eff}=-1-\frac{1}{3}\frac{d \ln h^2}{dx}$, where $x= \ln a$ and $h=H/H_0$. Using Eq. \eqref{eqb3}, we get
\begin{equation}\label{eqb10}
w_{eff}(z)=-1+\frac{(1-\zeta_1-\Omega_{\Lambda0})(1+z)^{3(1-\zeta_1-\nu)}}{\left[\frac{(\Omega_{\Lambda0}-\nu)}{(1-\zeta_1-\nu)}+\left(1-\frac{(\Omega_{\Lambda0}-\nu)}{(1-\zeta_1-\nu)}\right)(1+z)^{3(1-\zeta_1-\nu)}\right]}
\end{equation}
The present value of $w_{eff}$ at $z=0$ is given by
\begin{equation}\label{eqb11}
w_{eff}(z=0)=-1+(1-\zeta_1-\Omega_{\Lambda0}).
\end{equation}
We can observe that the model will accelerate provided $3w_{eff}(z=0)+1=-2+3(1-\zeta_1-\Omega_{\Lambda0})<0$.\\
\indent Let us discuss the behavior of the matter energy  density in this model as a function of scale factor (or redshift). Transforming the time derivative into derivative with respect to the scale factor, the  conservation equation \eqref{ce1} reduces to a differential equation for matter density,
\begin{equation}\label{eqb13}
\frac{d\rho_m}{da}+\frac{3(1-\nu)}{a}\rho_m=\frac{9(1-\nu)\zeta_1}{a}H^2,
\end{equation}
Using \eqref{eqb3} into \eqref{eqb13} and integrating, we find
\begin{align}\label{eqb14}
\rho_m&=\left(\rho_{m0}-\frac{3\zeta_1H_0^2(\Omega_{\Lambda 0}-\nu)}{(1-\zeta_1-\nu)}\right)a^{-3(1-\zeta_1-\nu)}\nonumber\\
&+\frac{3\zeta_1H_0^2(\Omega_{\Lambda 0}-\nu)}{(1-\zeta_1-\nu)},
\end{align}
where $\rho_{m0}=\rho_m(a=1)$ is the present matter density. Substituting Eq. \eqref{eqb4} in the above equation, one may obtain the explicit time evolution of the matter density if desired. It can be observed from Eq. \eqref{eqb14} that the matter density does not no longer evolve as $\rho_m=\rho_{m0}\;a^{-3}$. There is a correction in the exponent of the scale factor and some additional constant terms. This is due to the fact that matter is exchanging energy from vacuum and viscous term. We also note that as $t\rightarrow \infty$, $\rho_m=3\zeta_1H_0^2(\Omega_{\Lambda 0}-\nu)/(1-\zeta_1-\nu)$, i.e, matter density does not approach zero in infinitely far future due to viscosity. In the absence of viscous term, the matter density tends to zero as $t\rightarrow \infty$. The detail discussion on the evolutions of matter energy density and other cosmological parameters of this particular model is presented in Section VII.\\
\indent In the following Section, we constrain the parameters of this model by using the latest observational data sets and analyze the evolutions of all above discussed various cosmological parameters using the best-fit values. We compare the proposed model with the existing model through the stability criteria.
\section{Growth of perturbations}\label{perturbation}
In cosmic structure formation it is assumed that the present abundant structure of the Universe is developed through gravitational amplification of small density perturbations generated in its early evolution. In this section, we briefly discuss the linear perturbation within the framework of viscous fluid with varying $\Lambda(t)$. We refer the reader to Refs. \cite{sola2018,agvalent2018} for the detailed perturbation equations since here we have discussed some basic equations only. The differential equation for the matter density contrast  $\delta_m \equiv \delta \rho_{m}/ \rho_{m}$ for our model considered here can be approximated as follows \cite{decruz2023}:
\begin{equation}\label{stf}
\delta_{m}''+ \left( \frac{3}{a} + \frac{H'(a)}{H(a)} \right) {\delta_{m}'} - \frac{4\pi G \rho_m}{H^2(a)} \frac{\delta_{m}}{a^2} = 0
\end{equation}
where prime represents derivative with respect to the scale factor $a$.  The above second-order differential equation turns out to be accurate since the main effects come from the different expression of the Hubble function. We consider the Hubble function as obtained in {\it case III of Sect. IV}. Equation \eqref{stf} describes the smoothness of the matter perturbation in extended viscous $\Lambda(t)$ model.\\
\indent The linear growth rate of the density contrast, $f$, which is related to the peculiar velocity in the linear theory \cite{pebbles1993} is defined as
\begin{equation}\label{gp1}
f(a)=\frac{d \ln D_{m}(a)}{ d\ln a},
\end{equation}
where $D_m(a)=\delta_{m}(a)/\delta_{m}(a=1)$ is the linear growth function. The weighted linear growth rate, denoted by $f\sigma_8$, is the product of the growth rate $f(z)$, defined in \eqref{gp1}, and $\sigma_8(z)$. Here, $\sigma_8 $ is the root-mean-square fluctuation in spheres with radius $8 h^{-1}$ Mpc scales \cite{song2009,huterer2015}, and it is given by \cite{nesseris2008}
\begin{equation}\label{gp2}
\sigma_{8}(z)=\frac{\delta_m(z)}{\delta_m(z=0)}\sigma_8(z=0).
\end{equation}
Using \eqref{gp1} and \eqref{gp2}, the weighted linear growth rate is given by
\begin{equation}\label{gp3}
f\sigma_8(z)=-(1+z)\frac{\sigma_8(z=0)}{\delta_m(z=0)}\frac{d \delta_{m}}{dz}.
\end{equation}
\indent In what  follows, we perform the observational analysis of case III of Section IV to estimate the parameters of the model and analyse the evolution and dynamics of the model in detail.
\section{Data and methodology}
\label{sec:3}
In this section, we present the data and methodology used in this work.  We constrain the parameters of the $GR-\Lambda$CDM and $\zeta=\zeta_1 H$ with varying $\Lambda$ models using a large, robust and latest set of observational data which involve observations from: (i) distant type Ia supernovae (SNe Ia); (ii) a compilation of cosmic chronometer measurements of Hubble parameter $H(z)$ at different redshifts; (iii)  baryonic acoustic oscillations (BAO); and (iv) $f(z)\sigma_8(z)$ data. A brief description of each of datasets are as follows:
\subsection{Pantheon SNe Ia sample}
The most known and frequently used cosmological probe are distant type Ia supernovae (SNe Ia) which are used to understand the actual evolution of the Universe. A supernova explosion is an extremely luminous event, with its brightness being comparable with the brightness of its host galaxy \cite{scolnic2018}. We use the recent SNe Ia data points, the so-called Pantheon sample which includes 1048 data points of luminosity distance in the redshift range $0.01 <z <2.26$. Specifically, one could use the observed distance modulo, $\mu_{obs}$ , to constrain cosmological models. The Chi-squared function for SNe Ia is given by
\begin{equation}
\chi^2_{SNe\;Ia} =  \sum_{i=1}^{1048} \Delta \mu^{T} C^{-1} \Delta \mu,
\end{equation}
where $\Delta \mu = \mu_{obs} - \mu_{th}$. Here, $\mu_{obs}$ is the observational distance modulus of SNe Ia and is given as $\mu_{obs}=m_B-\mathcal{M}$, where $m_B$ is the observed peak magnitude in the rest frame of the $B$ band, $\mathcal{M}$ is the absolute B-band magnitude of a fiducial SNe Ia, which is taken as $-19.38$. The theoretical distance modulus $\mu_{th}$ is defined by
\begin{equation}
 \mu_{th}(z,{\bf p}) = 5 \log_{10}\left(\frac{D_{L}(z_{hel},z_{cmb})}{1 M pc} \right) + 25,
 \end{equation}
where $\bf{p}$ is the parameter space and  $D_{L}$ is the luminosity distance, which is given as $D_{L}(z_{hel},z_{cmb}) = (1 + z_{hel})r(z_{cmb})$. Here, $r(z_{cmb})$ is given by
\begin{equation}
r(z)= c H_0^{-1}\int_0^{z} \frac{ dz'}{E(z',{\bf p})},
\end{equation}
where $c$ is the speed of light, $E(z)\equiv H(z)/H_0$ is the dimensionless Hubble parameter, $z_{hel}$ and $z_{cmb}$ are heliocentric and CMB frame redshifts, respectively. Here, $C$ is the total covariance matrix which takes the form $C=D_{stat}+C_{sys}$, where the diagonal matrix $D_{stat}$ and covariant matrix $C_{sys}$ denote the statistical uncertainties and the systematic uncertainties.
\subsection{BAO measurements}
In this work, we have used six points of BAO data-sets from several surveys, which includes the Six Degree Field Galaxy Survey (6dFGS),the Sloan Digital Sky Survey (SDSS), and the LOWZ samples of the Baryon Oscillation Spectroscopic Survey(BOSS)\cite{Blake,Percival,Giostri}.\\
\indent The dilation scale $D_{v}(z)$ introduced in \cite{Eisenstein} is given by
\begin{equation}
D_v(z)=\left( \frac{d_A^2(z)z}{H(z)}\right)^{1/3}
\end{equation}
Here, $d_A(z)$ is the comoving angular diameter distance and is defined as\\
\begin{equation}
d_A(z)=\int_{0}^{z}\frac{dy}{H(y)'},
\end{equation}
Now, the corresponding Chi-squared function for the BAO analysis is given by
\begin{equation}
\chi_{BAO}^2=A^T C^{-1}_{BAO} A,
\end{equation}
where $A$ depend on the considered survey and $C^{-1}_{BAO}$ is the inverse of the covariance matrix \cite{Giostri}.
\subsection{$H(z)$ data}
The cosmic chronometer (CC) data, which is determined by using the most massive and passively evolving galaxies based on the `galaxy differential age' method, are model independent (see, Ref.\cite{moresco2022} for detail). In our analysis, we use 32 CC data points of the Hubble parameter measured by differential age technique \cite{moresco2022} between the redshift range $0.07 \leq z \leq 1.965 $.  The Chi-squared function for $H(z)$ is given by
\begin{equation}
\chi_{H(z)}^2=\sum_{i=1}^{32}\frac{[H(z_i,{\bf p})-H_{obs}(z_i)]^2}{\sigma^2_{H(z_i)}}
\end{equation}
where $H(z_i,{\bf p})$ represents the theoretical values of Hubble parameter with model parameters, $H_{obs}(z_i)$ is the observed values of Hubble parameter and $\sigma_i$ represents the standard deviation measurement uncertainty in $H_{obs}(z_i)$.
\subsection{ $f(z)\sigma_8(z)$ data}
In Section IV, we have mainly discussed the background evolution of the growth perturbations and defined the weighted linear growth rate by Eq. \eqref{gp3}. To make more complete discussion on viscous $\Lambda(t)$ model in perturbation evolution, we focus on an observable quantity of $f(z)\sigma_8(z)$. We use 18 data points of ``Gold -17" compilation of robust and independent measurements of weighted linear growth $f(z)\sigma_8(z)$ obtained by various galaxy surveys as complied in Table III of Ref. \cite{nesseris2017}. In order to compare the observational data set with that of the predicted by our model, we define the Chi-square function as
\begin{equation}\label{gp4}
\chi^2_{(f\sigma_8)} = \sum_{i=1}^{18}\frac{[f\sigma^{the}_8(z_i,{\bf p})-f\sigma^{obs}_8(z_i)]^2}{\sigma^2_{f\sigma_8(z_i)}},
\end{equation}
where $f\sigma^{the}_8(z_i,{\bf p})$ is the theoretical value computed by Eq.\eqref{gp3} and $f\sigma^{obs}_8(z_i)$ is the observed data \cite{nesseris2017}.
\begin{figure*}
\begin{minipage}[t]{0.32\linewidth}
\includegraphics[width=\linewidth]{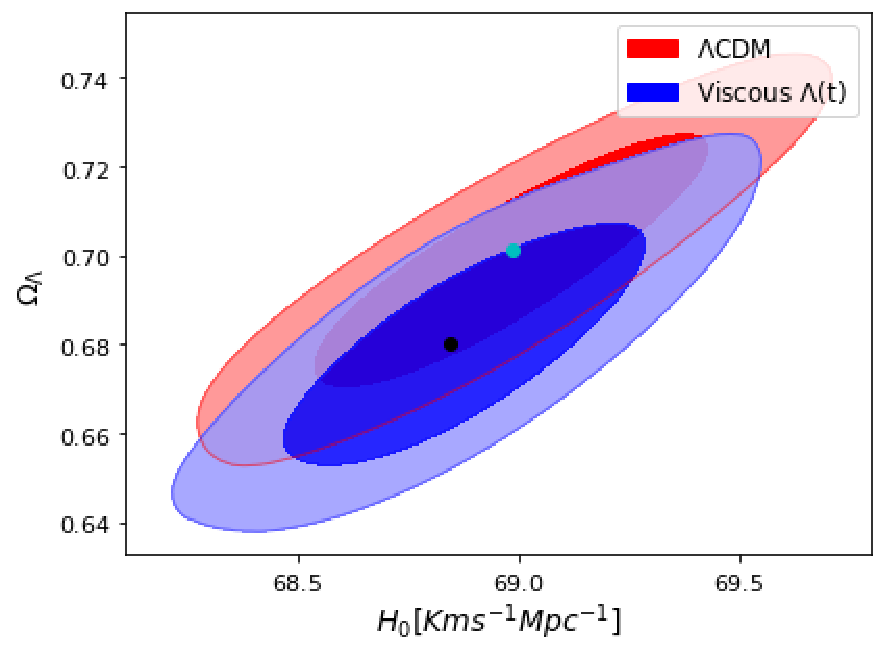}
\end{minipage}
\begin{minipage}[t]{0.32\linewidth}
\includegraphics[width=\linewidth]{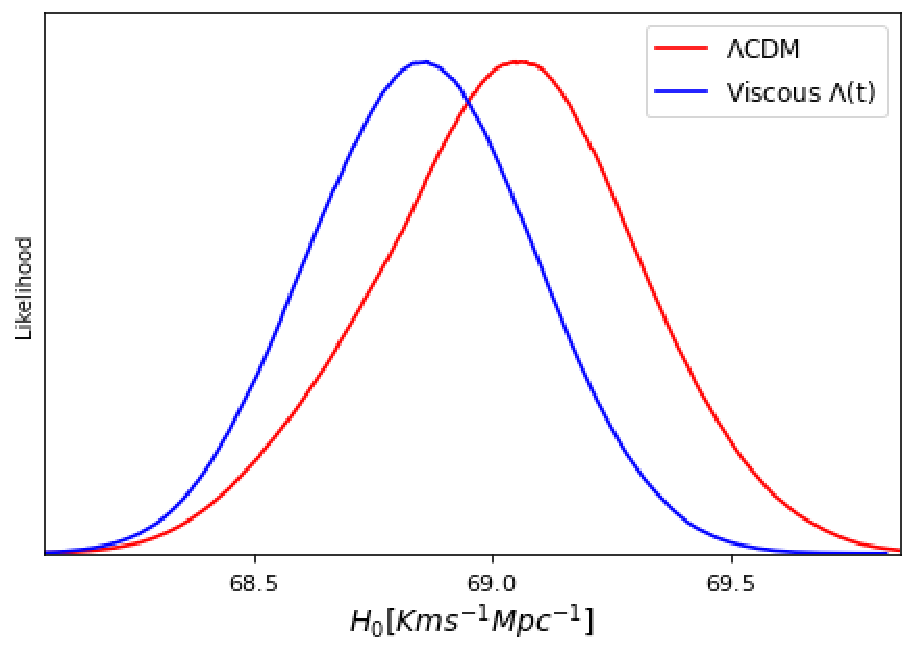}
\end{minipage}
\begin{minipage}[t]{0.32\linewidth}
\includegraphics[width=\linewidth]{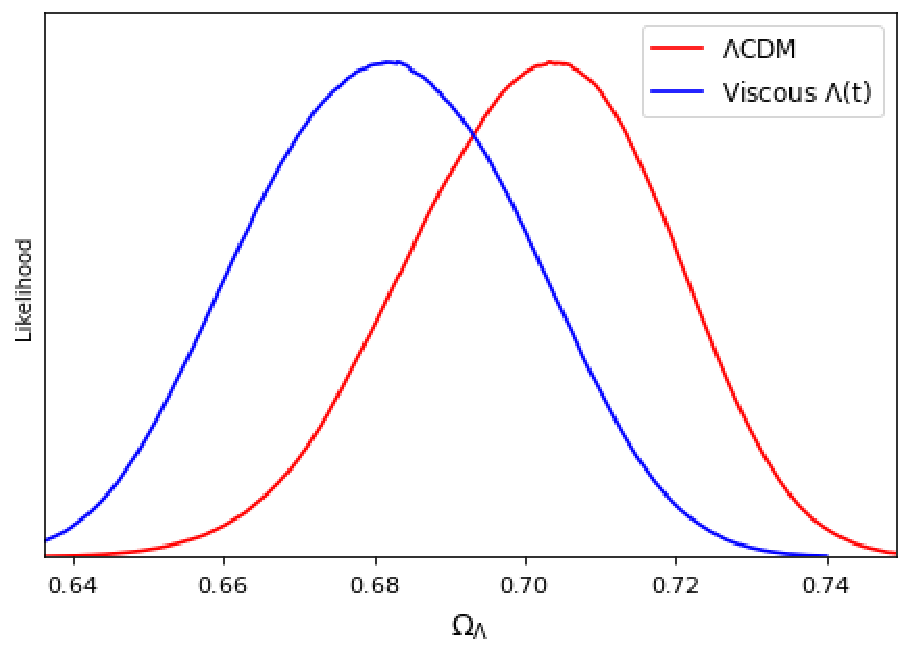}
\end{minipage}
\caption{Two-dimensional confidence contours of the $H_0 - \Omega_{\Lambda}$  and one dimensional posterior distributions of $H_0$, $\Omega_{\Lambda}$ for the $\Lambda$CDM and viscous $\Lambda(t)$ models using $``BASE"$ data. The green and black dot on the contour represents the best fit value of $\Lambda$CDM and viscous $\Lambda(t)$ models respectively.}
\end{figure*}\label{fig:4}
\begin{figure*}
\begin{minipage}[t]{0.32\linewidth}
\includegraphics[width=\linewidth]{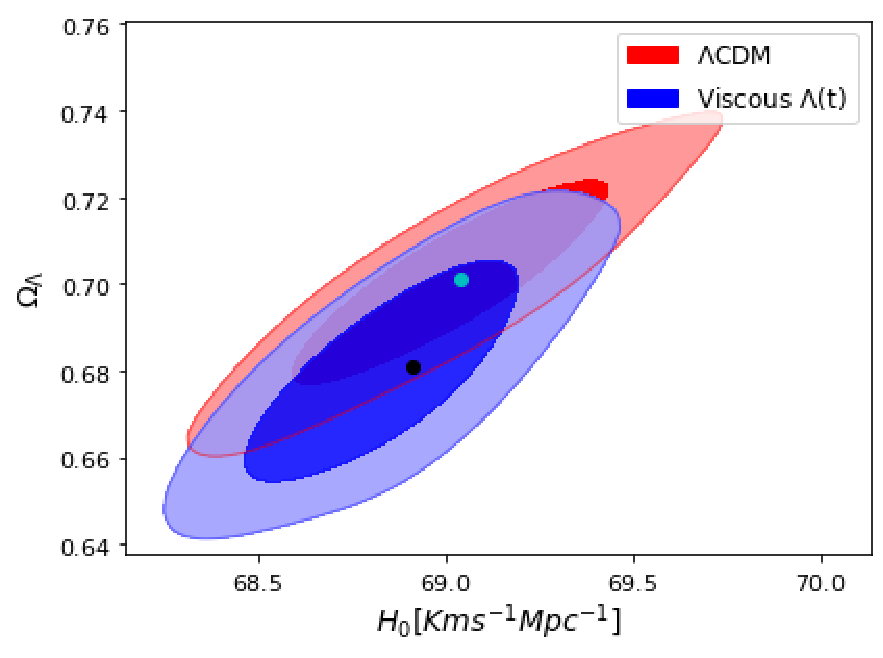}
\end{minipage}
\begin{minipage}[t]{0.32\linewidth}
\includegraphics[width=\linewidth]{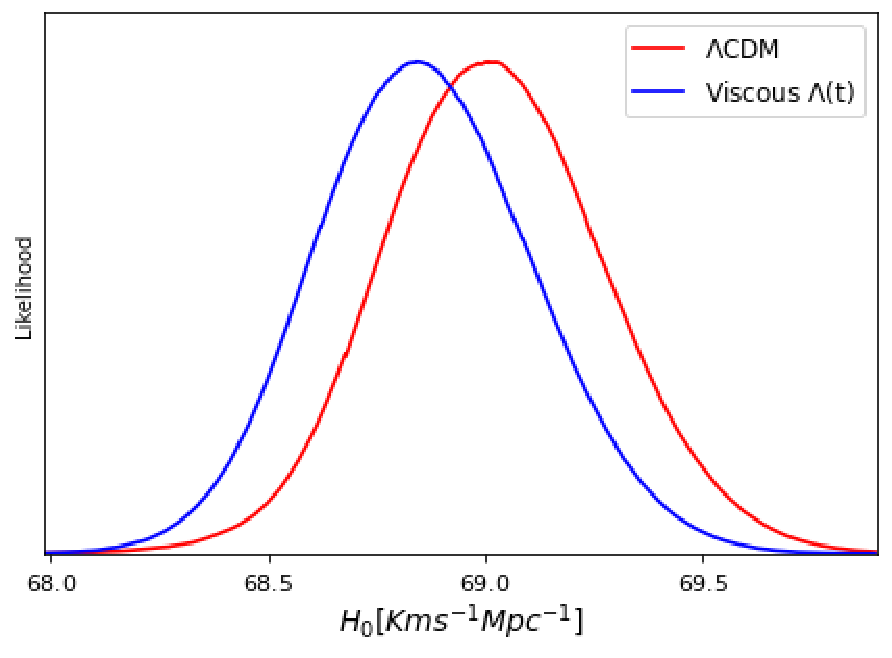}
\end{minipage}
\begin{minipage}[t]{0.32\linewidth}
\includegraphics[width=\linewidth]{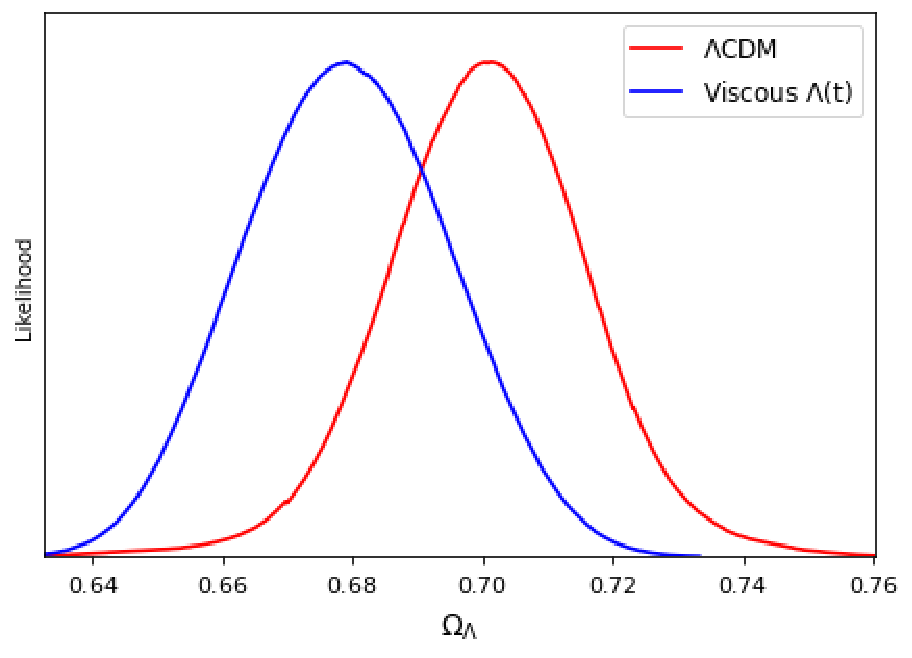}
\end{minipage}
\caption{Two-dimensional confidence contours of the $H_0 - \Omega_{\Lambda}$  and one dimensional posterior distributions of $H_0$, $\Omega_{\Lambda}$ for the $\Lambda$CDM and viscous $\Lambda(t)$ models using $``+CC"$ data. The green and black dot on the contour represents the best fit value of $\Lambda$CDM and viscous $\Lambda(t)$ models respectively.}
\end{figure*}\label{fig:5}

\begin{figure*}
\begin{minipage}[t]{0.32\linewidth}
\includegraphics[width=\linewidth]{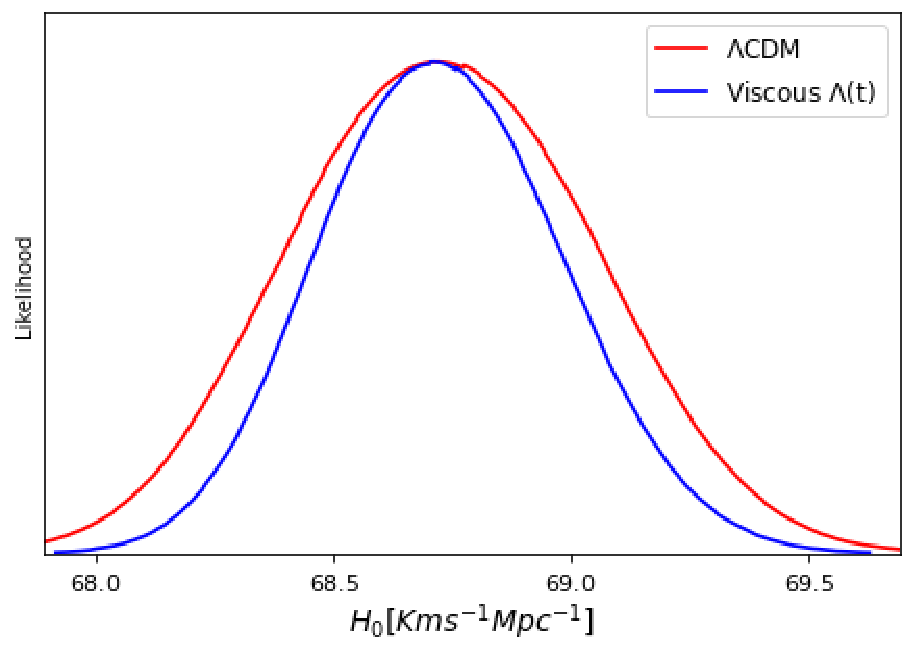}
\end{minipage}
\begin{minipage}[t]{0.32\linewidth}
\includegraphics[width=\linewidth]{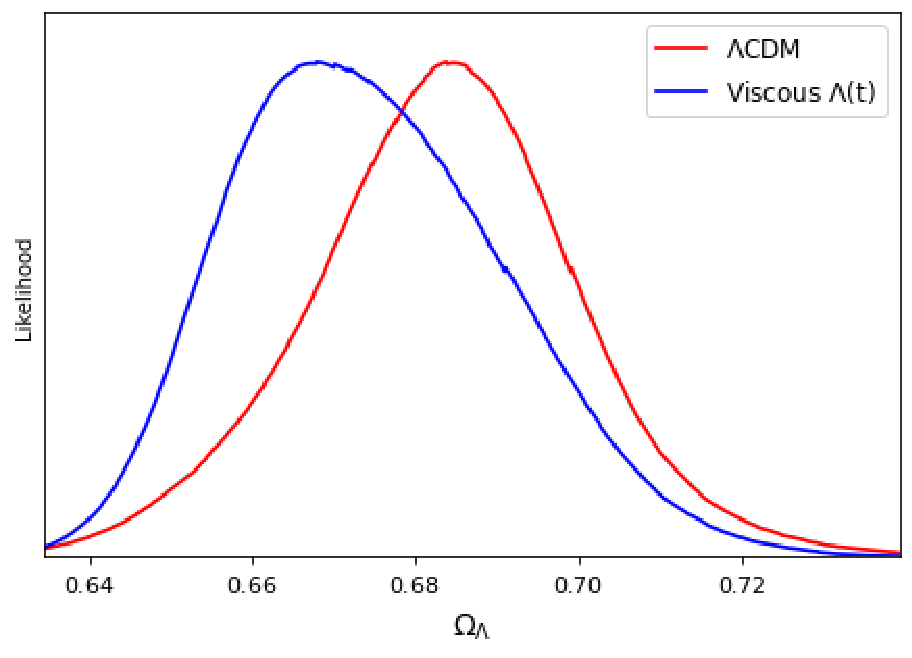}
\end{minipage}
\begin{minipage}[t]{0.32\linewidth}
\includegraphics[width=\linewidth]{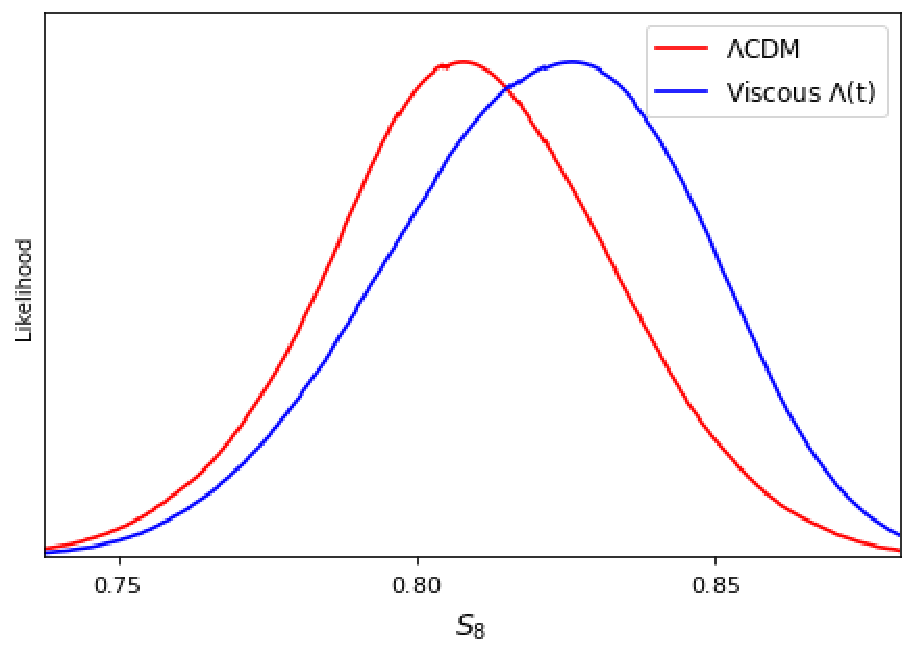}
\end{minipage}

\begin{minipage}[t]{0.32\linewidth}
\includegraphics[width=\linewidth]{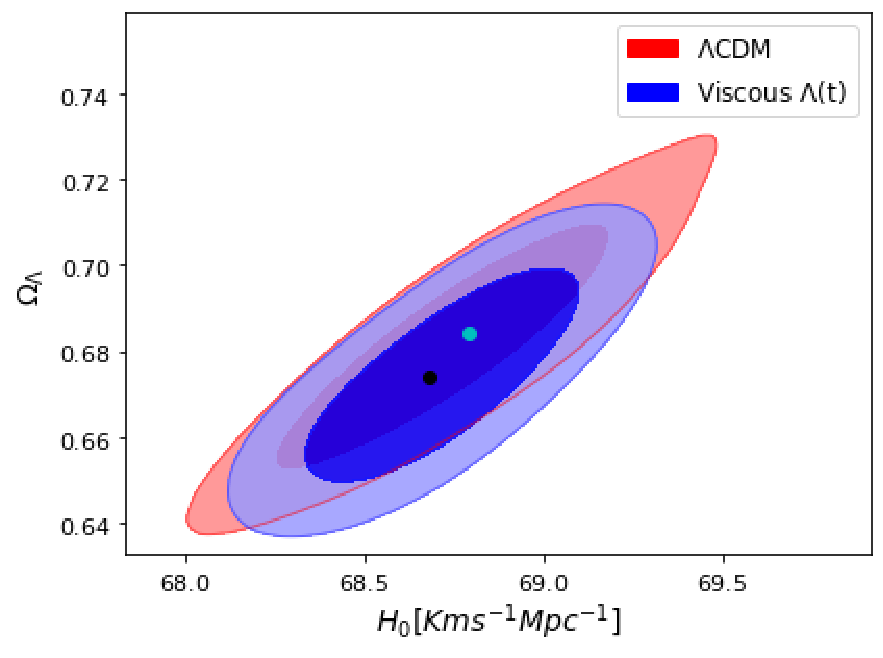}
\end{minipage}
\begin{minipage}[t]{0.32\linewidth}
\includegraphics[width=\linewidth]{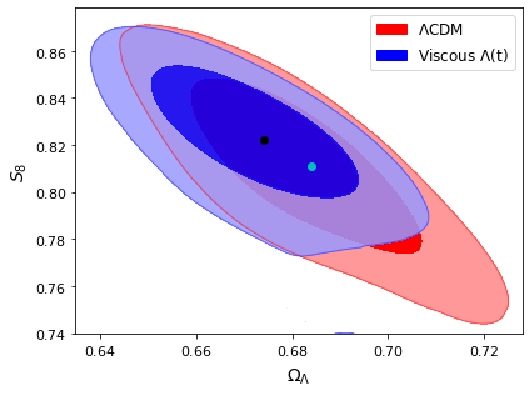}
\end{minipage}
\begin{minipage}[t]{0.32\linewidth}
\includegraphics[width=\linewidth]{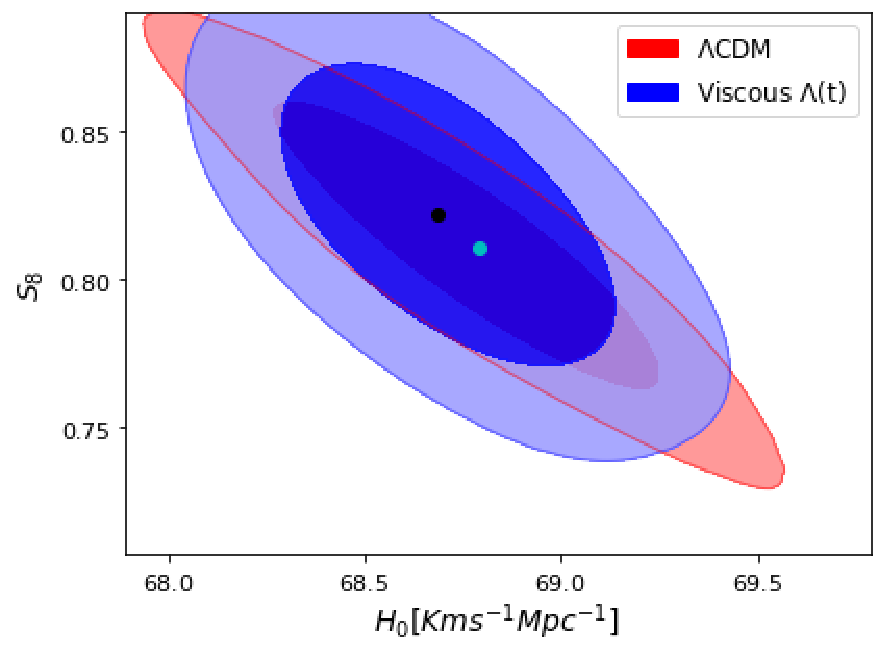}
\end{minipage}
\caption{Two-dimensional confidence contours of  $H_0 - \Omega_{\Lambda}$, $\Omega_{\Lambda} - S_8$ and $H_0 - S_8$ and one-dimensional posterior distributions of $H_0$, $\Omega_{\Lambda}$ and $S_8$ for the $\Lambda$CDM and viscous $\Lambda(t)$ models using $``+f\sigma_8"$ data. The green and black dot on the contour represents the best fit value of $\Lambda$CDM and viscous $\Lambda(t)$ models respectively.}
\end{figure*}\label{fig:6}
\indent Using the observational data as discussed above, we use the Markov Chain Monte Carlo (MCMC) method by employing EMCEE python package \cite{foreman2013} to explore the parameter spaces of viscous model with decaying vacuum density as discussed in part B of Sect.III by utilizing different combinations of data sets. The combinations are as follows:\\
\begin{itemize}
\item BASE: The combination of two datasets $SNe\;Ia+BAO$ is termed as ``BASE", whose the joint $\chi^2$ function is defined as $\chi^2_{tot}=\chi^2_{SNe\;Ia}+\chi^2_{BAO}$.
\item \textbf{$+CC$}: We combine $CC$ data  to the BASE, where  $\chi^2_{tot}=\chi^2_{SNe\;Ia}+\chi^2_{BAO}+\chi^2_{H(z)}$
\item \textbf{$+f\sigma_8(z)$}: The BASE data is complemented with $CC$ and $f\sigma_8$, where $\chi^2_{tot}=\chi^2_{SNe\;Ia}+\chi^2_{BAO}+\chi^2_{H(z)}+\chi^2_{f\sigma_8}$.
\end{itemize}
We consider the $\Lambda$CDM model as a reference model and its parameters are also constrained with the above sets of data.
\begin{table}
\caption{Constraints on parameters of $\Lambda$CDM for different set of observation data. Here ``BASE" denotes ``SNe Ia+BAO"}
\begin{tabular}{ |l||c|c|c|}
\hline
\multicolumn{4}{|c|}{$\Lambda$CDM} \\
\hline
Parameter & BASE & +CC & +$f\sigma_8$ \\
\hline
$H_0$ & $68.987^{+0.263}_{-0.276}$ & $69.001^{+0.238}_{-0.223}$  &  $68.793^{+0.193}_{-0.221}$  \\

$\Omega_\Lambda$ & $0.701^{+0.013}_{-0.020}$ & $0.699^{+0.016}_{-0.015}$  &  $0.684^{+0.015}_{-0.014}$  \\	

$\sigma_8$ & $-$ & $-$  &  $0.794^{+0.014}_{-0.015}$  \\

$S_8$ & $-$ & $-$  &  $0.811^{+0.022}_{-0.022}$  \\

$z_{tr}$ & $0.670^{+0.038}_{-0.038}$  & $0.674^{+0.035}_{-0.035}$  & $0.625^{+0.041}_{-0.041}$\\

$q_{0}$ & $-0.549^{+0.020}_{-0.023}$  & $-0.551^{+0.020}_{-0.020}$  & $-0.523^{+0.025}_{-0.025}$  \\

$w_{0}$ & $-0.699^{+0.013}_{-0.015}$ & $-0.701^{+0.013}_{-0.013}$ & $-0.682^{+0.017}_{-0.017}$  \\

$t_0(Gyr)$ & $13.73^{+0.017}_{-0.017}$  & $13.69^{+0.015}_{-0.015}$ & $13.54^{+0.013}_{-0.013}$ \\
\hline
\end{tabular}
\label{table:values}
\end{table}
\begin{table}
\caption{Constraints on parameters of viscous $\Lambda$(t) model using different set of observation data.}
\begin{tabular}{ |l||c|c|c|}
\hline
\multicolumn{4}{|c|}{Viscous $\Lambda$(t)} \\
\hline
Parameter & BASE & +CC & +$f\sigma_8$ \\
\hline
$H_0$  & $68.843^{+0.274}_{-0.238}$ & $68.913^{+0.262}_{-0.261}$  &  $68.684^{+0.259}_{-0.241}$  \\

$\Omega_\Lambda$ & $0.680^{+0.018}_{-0.020}$ & $0.684^{+0.013}_{-0.020}$  &  $0.674^{+0.012}_{-0.016}$  \\

$\zeta_1$ & $0.006^{+0.007}_{-0.004}$ & $0.006^{+0.008}_{-0.004}$  &  $0.003^{+0.005}_{-0.002}$  \\

$\nu$ & $0.004^{+0.003}_{-0.003}$ & $0.003^{+0.004}_{-0.002}$  &  $0.003^{+0.004}_{-0.002}$  \\	

$\sigma_8$ & $-$ & $-$  &  $0.790^{+0.008}_{-0.010}$  \\

$S_8$ & $-$ & $-$  &  $0.822^{+0.019}_{-0.019}$  \\

$z_{tr}$  & $0.664^{+0.031}_{-0.042}$ & $0.665^{+0.031}_{-0.037}$ & $0.626^{+0.028}_{-0.038}$\\

$q_{0}$  & $-0.533^{+0.025}_{-0.020}$ & $-0.535^{+0.023}_{-0.020}$ & $-0.516^{+0.022}_{-0.017}$ \\

$w_{0}$  & $-0.689^{+0.017}_{-0.013}$ & $-0.690^{+0.015}_{-0.013}$ & $-0.677^{+0.014}_{-0.011}$ \\

$t_0(Gyr)$ & $13.52^{+0.019}_{-0.019}$ & $13.48^{+0.017}_{-0.017}$  & $13.47^{+0.013}_{-0.015}$\\
\hline
\end{tabular}
\label{table:values1}
\end{table}
\section{Results and Discussion}\label{result}
In this section, we present the main results obtained through the observational data on the viscous $\Lambda(t)$ model of the form $\zeta=\zeta_1 H$ with $\Lambda=c_0+3\nu H^2$ (Refers to case III of Sect.IV). We also present the cosmological observation for $\Lambda$CDM model using the three combination of datasets. The viscous $\Lambda(t)$ model has 4 free parameter spaces $\{H_0, \Omega_\Lambda, \zeta_1, \nu\}$, where as $\Lambda$CDM has 2 free parameters $\{H_0, \Omega_{\Lambda}\}$. We calculate the best-fit values by minimizing the combination of $\chi^2$ function for above defined data sets. We also provide the fitting values of the $\Lambda$CDM for comparison with the viscous $\Lambda(t)$ model. The constraints of the statistical study are presented in Tables I and II. Figures 4-6 show the $1\sigma (68.3 \%)$ and $2\sigma (95.4\%)$ confidence level (CL) contours with marginalized likelihood distributions for the cosmological parameters of $\Lambda$CDM and viscous $\Lambda(t)$ models considering combination of different datasets, respectively. It is observed from Tables I and II that the constraints on the parameter spaces of $\Lambda$CDM and viscous with $\Lambda(t)$ are nearly the same.  \\
\begin{figure*}
\begin{minipage}[t]{0.48\linewidth}
\includegraphics[width=\linewidth]{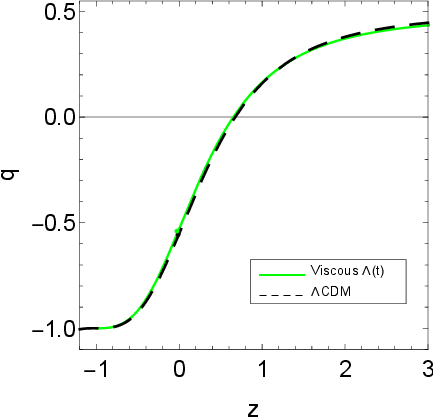}
\caption{The redshift evolution of the deceleration parameter for viscous $\Lambda(t)$ using $``BASE"$ dataset. The evolution of deceleration parameter in the standard $\Lambda$CDM model is also shown as the dashed curve. A dot denotes the current value of $q$ (hence $q_0$).}\label{fig:7}
\end{minipage}
\hfill
\begin{minipage}[t]{0.48\linewidth}
\includegraphics[width=\linewidth]{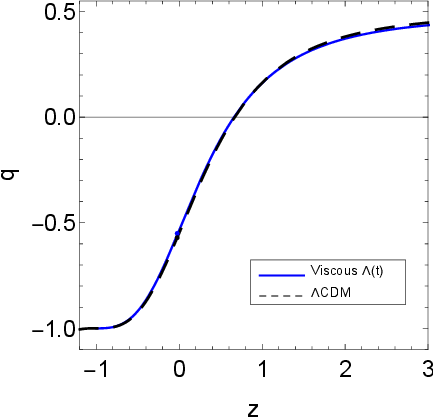}
\caption{The redshift evolution of the deceleration parameter for viscous $\Lambda(t)$ using $``+CC"$ dataset. The evolution of deceleration parameter in the standard $\Lambda$CDM model is also shown as the dashed curve. A dot denotes the current value of $q$ (hence $q_0$).}\label{fig:8}
\end{minipage}
\end{figure*}
\begin{figure*}
\begin{minipage}[t]{0.48\linewidth}
\includegraphics[width=\linewidth]{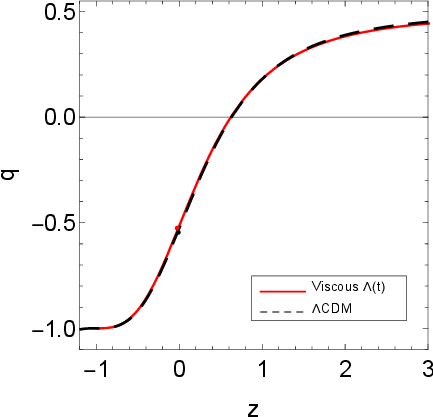}
\caption{The redshift evolution of the deceleration parameter for viscous $\Lambda(t)$ using $``+f\sigma_8"$ dataset. The evolution of deceleration parameter in the standard $\Lambda$CDM model is also shown as the dashed curve. A dot denotes the current value of $q$ (hence $q_0$).} \label{fig:9}
\end{minipage}
\hfill
\begin{minipage}[t]{0.49\linewidth}
\includegraphics[width=\linewidth]{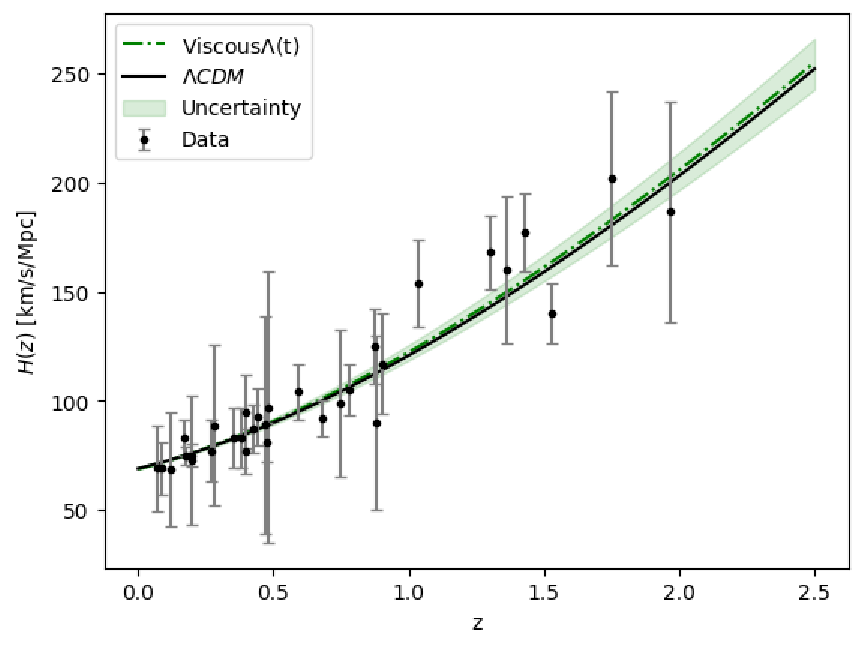}
\caption{Best fits using $``BASE"$ data set over $H(z)$ data for viscous $\Lambda(t)$ (green dot-dashed line) and $\Lambda$CDM (black  solid line) are shown. The grey points with uncertainty bars correspond to the 32 $CC$ sample.} \label{fig:10}
\end{minipage}
\end{figure*}
\begin{figure*}
\begin{minipage}[t]{0.49\linewidth}
\includegraphics[width=\linewidth]{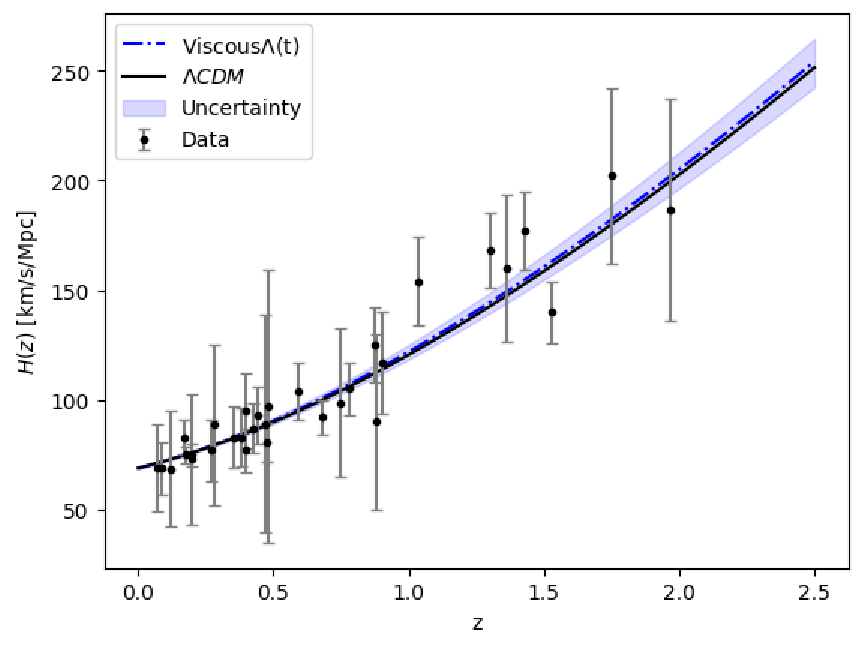}
\caption{Best fits using $``+CC"$ data set over $H(z)$ data for viscous $\Lambda(t)$ (blue dot-dashed line) and $\Lambda$CDM (black solid line) are shown. The grey points with uncertainty bars correspond to the 32 $CC$ sample.} \label{fig:11}
\end{minipage}
\hfill
\begin{minipage}[t]{0.49\linewidth}
\includegraphics[width=\linewidth]{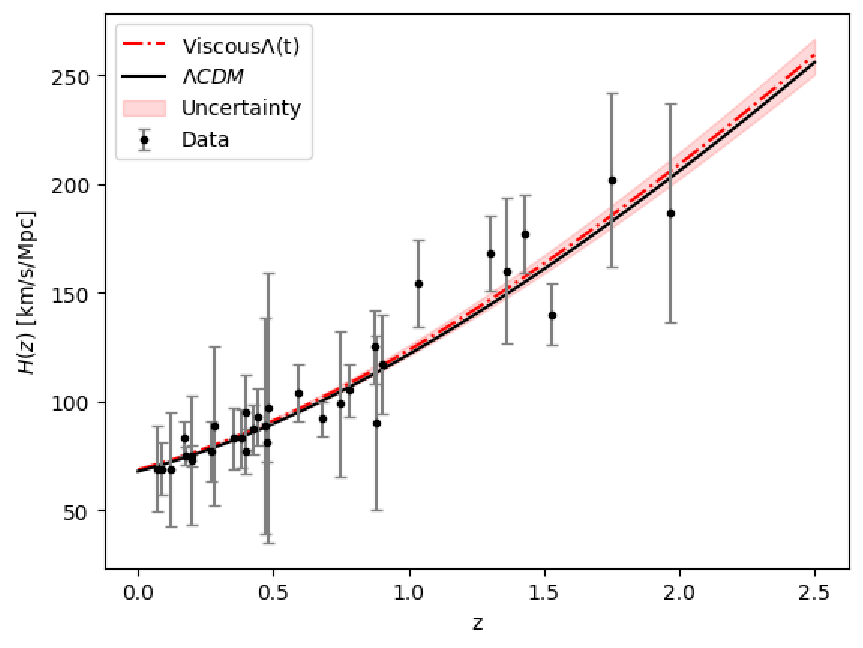}
\caption{Best fits using $``+f\sigma_8"$ data set over $H(z)$ data for viscous $\Lambda(t)$ (red dot-dashed line) and $\Lambda$CDM (black solid line) are shown. The grey points with uncertainty bars correspond to the 32 $CC$ sample.} \label{fig:12}
\end{minipage}
\end{figure*}
\indent Using best-fit values of parameters obtained from $BASE$, $+CC$ and $+f \sigma_8$ data into Eq.\eqref{eqb7}, the evolutions of the deceleration parameter with respect to the redshift are shown in Figs.7-9 for viscous $\Lambda(t)$ model along with the $\Lambda$CDM model. It is observed that with each data set $q(z)$ varies from positive to negative and show the similar trajectory  that is comparable to the $\Lambda$CDM model. Thus, both the models depict a transition from the early decelerated phase to the late-time accelerated phase. Further, $q(z)$ approaches to $-1$ in late-time of evolution. Thus, the models successfully generate late-time cosmic acceleration along with a decelerated expansion in the past. Figures 7-9 show that the transition from decelerated to accelerated phase take place at redshift $z_{tr}=0.664^{+0.031}_{-0.042}$ with $BASE$ data, $z_{tr}=0.665^{+0.031}_{-0.037}$ with $+CC$ data and  $z_{tr}=0.626^{+0.028}_{-0.037}$ with  $+f \sigma_8$ data. The datasets $BASE$, $+CC$ and $+f \sigma_8$ yield the present deceleration parameter $q_0$ as $-0.533^{+0.025}_{-0.020}$, $-0.535^{+0.023}_{-0.020}$ and $-0.516^{+0.022}_{-0.017}$ respectively (cf. Table II). The present values of $z_{tr}$ and $q_0$ are very close and thus are in good agreement to $\Lambda$CDM as presented in Table I.   \\
\indent The evolutions of the Hubble parameter $H(z)$ of viscous $\Lambda(t)$ model with respect to the redshift are shown in Figs. 10-12. Throughout the expansion, viscous $\Lambda(t)$ is coinciding with the $\Lambda$CDM model and the model paths cover majority of the dataset with the error bar of Hubble parameter, indicating that the viscous $\Lambda(t)$ agrees well with the $\Lambda$CDM model for all the three combination of datasets. In the considered cosmological scenario, the present age of the Universe are found to be $t_0 \approx 13.52 \;Gyr$ , $t_0\approx 13.48\;Gyr$ and $t_0\approx 13.47\;Gyr$ respectively as presented in Table II. The ages thus obtained are very much compatible with that obtained from the $\Lambda$CDM model with the same datasets (cf.Table I). \\
\begin{figure*}
\begin{minipage}[t]{0.49\linewidth}
\includegraphics[width=\linewidth]{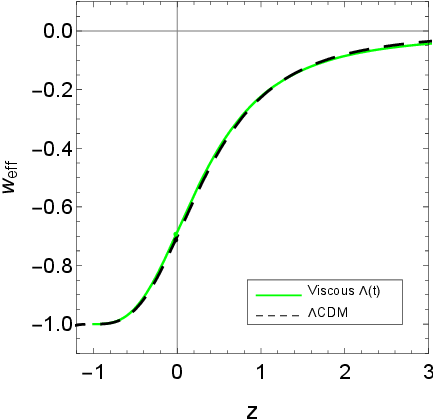}
\caption{Effective EoS parameter as a function of redshift $z$ for viscous $\Lambda(t)$ using $``BASE"$ dataset. The evolution of EoS parameter in the standard $\Lambda$CDM model is also represented as the dashed curve. A dot denotes the present value of the EoS parameter.} \label{fig:13}
\end{minipage}
\hfill
\begin{minipage}[t]{0.49\linewidth}
\includegraphics[width=\linewidth]{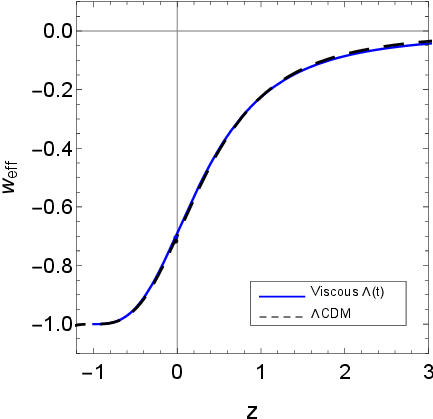}
\caption{Effective EoS parameter as a function of redshift $z$ for viscous $\Lambda(t)$ using $``+CC"$ dataset. The evolution of EoS parameter in the standard $\Lambda$CDM model is also represented as the dashed curve. A dot denotes the present value of the EoS parameter.} \label{fig:14}
\end{minipage}
\end{figure*}
\begin{figure*}
\begin{minipage}[t]{0.49\linewidth}
\includegraphics[width=\linewidth]{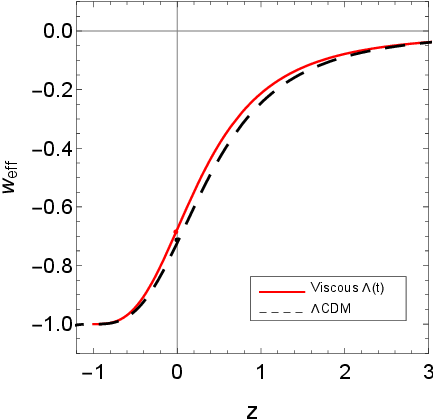}
\caption{Effective EoS parameter as a function of redshift $z$ for viscous $\Lambda(t)$ using $``+f\sigma_8"$ dataset. The evolution of EoS parameter in the standard $\Lambda$CDM model is also represented as the dashed curve. A dot denotes the present value of the EoS parameter.} \label{fig:15}
\end{minipage}
\hfill
\begin{minipage}[t]{0.49\linewidth}
\includegraphics[width=\linewidth]{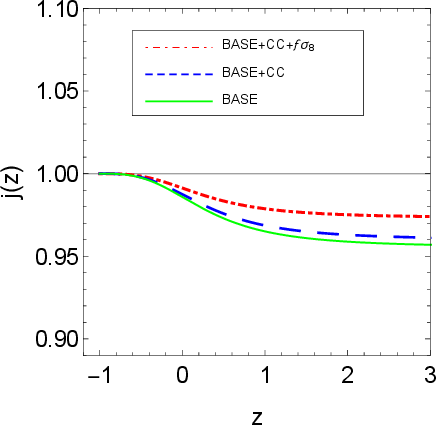}
\caption{Jerk parameter $j(z)$ with redshift $z$ using best-fit values of parameters for viscous $\Lambda(t)$ model. The horizontal line represents the $\Lambda$CDM model.} \label{fig:16}
\end{minipage}
\end{figure*}
\begin{table*}
\begin{ruledtabular}
\caption{Values of Chi-squared, reduced Chi-squared, AIC and BIC of $\Lambda$CDM and viscous $\Lambda(t)$ models. The $\Lambda$CDM model is considered as reference model to calculate the $\Delta$AIC and $\Delta$BIC.}
\begin{tabular}{|l|c|c|c|c|c|c|}
Values &  \multicolumn{2}{c|}{BASE} & \multicolumn{2}{c|}{+CC} & \multicolumn{2}{c|}{+$f\sigma_8$} \\
\cline{2-7}
& $\Lambda$CDM & viscous $\Lambda$(t) & $\Lambda$CDM & viscous $\Lambda$(t) & $\Lambda$CDM & viscous $\Lambda$(t)\\
\hline
$\chi^2$ & $518.017$ & $515.074$ & $525.457$ & $522.390$ & $842.630$ & $831.112$\\

$d$ & $2$ & $4$ & $2$ & $4$ & $2$ & $4$\\

$N$ & $1054$ & $1054$ & $1086$ & $1086$ & $1104$ & $1104$\\

$\chi^2_{red}$ & $0.492$ & $0.498$ & $0.484$ & $0.481$ & $0.764$ & $0.755$ \\

$AIC$ & $522.028$ & $523.055$ & $529.468$ & $530.427$ & $846.641$ & $839.112$\\

$BIC$ & $531.938$ & $542.915$ & $539.438$ & $550.351$ & $856.643$ & $859.139$\\

$\Delta$AIC & $-$ & $1.026$ & $-$ & $0.959$ & $-$ & $-7.492$\\

$\Delta$BIC & $-$ & $10.977$ & $-$ & $10.913$ & $-$ & $2.496$\\

\end{tabular}
\end{ruledtabular}
\label{table:values3}
\end{table*}
\indent Using the best-fit values of parameters in Eq. \eqref{eqb10}, the evolutions of the effective EoS parameter $w_{eff}$ are shown in Figs.13-15. We conclude that for large redshifts, $w_{eff}$ has small negative value $w_{eff}>-1/3$ and in future the model asymptotically approaches to $w_{eff}=-1$. The trajectory of $w_{eff}$ for $BASE$ and $+CC$ datasets coincides with the evolution of $\Lambda$CDM model. However, it slightly varies with the best-fit values obtained through $+f\sigma_8(z)$ data points. It can be observed that the viscous $\Lambda$(t) model behaves like a quintessence in early time and cosmological constant in late-time. The present values of $w_{eff}$ are found to be  $-0.689^{+0.017}_{-0.013}$, $-0.690^{+0.015}_{-0.013}$ and $-0.677^{+0.014}_{-0.011}$ with $BASE$, $+CC$ and $+f\sigma_8$ datasets respectively, which are very close to the current value of $\Lambda$CDM model as presented in Table I. \\
\indent From Tables I and II, let us discuss the present value $H_0$ of Hubble parameter in case of viscous $\Lambda$(t) and $\Lambda$CDM models. The viscous $\Lambda(t)$ model gives  $H_0 = 68.843_{-0.238}^{+0.274}$ km/s/Mpc with $BASE$ data, the $+CC$ data gives $H_0 = 68.913^{+0.262}_{-0.261}$ km/s/Mpc  and, finally, the $+f\sigma_8$ renders the present value: $H_0 = 68.684_{-0.241}^{+0.259}$ km/s/Mpc. Recently, the local measurement $H_0 = 73.04 \pm 1.04$ km/s/Mpc from Riess et al.\cite{riess2022} exhibits a strong tension with the Planck 2018 release $H_0=67.4 \pm 0.5$ km/s/Mpc \cite{agh2020} at the $4.89 \sigma$ confidence level. The residual tensions of our fitting results with respect to the latest local measurement $H_0 = 73.04 \pm 1.04$ km/s/Mpc \cite{riess2022} are $3.92 \sigma$, $3.85 \sigma$ and $4.07 \sigma$ respectively. \\
\indent Let us focus on $\sigma_8$ and $S_8$ which play very relevant role in structure formation. The best-fit values of these parameters for $\Lambda$CDM and viscous $\Lambda(t)$ models using $BASE+CC+f\sigma_8$ data are reported in Tables I and II, respectively. We can read off $\sigma_8$ = $0.794^{+0.014}_{-0.015}$ for $\Lambda$CDM model (cf.Table I), whereas the viscous $\Lambda(t)$ model prediction is $\sigma_8$ = $0.790^{+0.008}_{-0.010}$ (cf. Table II). This is a very good result, which can be rephrased in terms of the fitting value of the related LSS observable $S_8 = \sigma_8 \sqrt{(1-\Omega_\Lambda)/0.3}$ quoted in the Tables I and II: $S_8 = 0.811 \pm 0.022$ for $\Lambda$CDM and $S_8 = 0.822 \pm 0.019$ for viscous $\Lambda$(t) model. The values of $\sigma_8$ and $S_8$ for viscous $\Lambda(t)$ model is compatible for $1\sigma$ confidence level with $\Lambda$CDM. Our result predicts that the tensions in $\sigma_8$ and $S_8$ are reduced to $0.23\sigma$ and $-0.38\sigma$, respectively. The behavior of $f(z)\sigma_8(z)$ as a function of redshift is plotted in Fig.17. We can see that the evolution of $f\sigma_8$ for both viscous $\Lambda(t)$ and $\Lambda$CDM models are consistent with the observational data points.\\
\indent Table III presents the $\chi^2$ and reduced $\chi^2$ of $\Lambda$CDM and viscous $\Lambda(t)$ models, respectively for the used datasets. To compute reduced $\chi^2$, denoted as $\chi^2_{red}$, we use $\chi^2_{red}$ = $\chi^2_{min}/(N-d)$, where $N$ is the total number of data points and $d$ is the total number of fitted parameters, which differs for the various models. It should be noted that when a model is fitted to data, a value of $\chi^2_{red}<1$ is regarded as the best fit, whereas a value of $\chi^2_{red}>1$ is regarded as a poor fit. In our observations, we have used $N=1054$ data points for BASE (SNIa and BAO), $N=1086$ data points for BASE+CC and $N=1104$ data points for BASE+CC+$f \sigma_{8}$. The number of free parameters of viscous $\Lambda(t)$ is $d=4$ where as for $\Lambda$CDM it is $d=2$. Using these information, the $\chi^2_{red}$ for both the models are given in Table III. It can be observed that the value of $\chi^2_{red}$ is less than unity with every data sets for both the models which show that the both models are in a very good fit with these observational data sets and the observed data are consistent with the considered models.\\
\indent Using the three combination of data sets, we are also interested in investigating the cosmographical aspects of the models, such as jerk parameter, which is defined as
\begin{equation}
j=\frac{\dddot{a(t)}}{aH^3}=q(2q+1)+(1+z)\frac{dq}{dz}.
\end{equation}
\indent The jerk parameter which is a dimensionless third derivative of the scale factor, can provide us the simplest approach to search for departures from the $\Lambda$CDM model. It is noted that for $\Lambda$CDM model, $j=1$(const.) always. Thus, any deviation from $j=1$ would favor a non-$\Lambda$CDM model. In contrast to deceleration parameter which has negative values indicating accelerating Universe, the positive values of the jerk parameter show an accelerating rate of expansion. In Fig. 16, the evolutions of jerk parameter are shown for $\Lambda$CDM and viscous $\Lambda(t)$ models using the best-fit values of parameters obtained from three combination of datasets. It is obvious from the figure that this parameter remains positive and less than unity in past, and eventually tends to unity in late-time. Thus, the jerk parameter deviates in early time but it attains the same value as $\Lambda$CDM in late-time.\\
\indent Using the best-fit values of parameters from Table II in Eq.\eqref{eqb14}, we plot the matter energy density as a function of redshift for different combinations of datasets in Fig.18. It is observed that the matter density was too large at the beginning of the cosmic evolution. As $z\rightarrow -1$, the matter energy density tends to a finite value for all combinations of datasets.
\begin{figure*}[t]
\begin{minipage}[t]{0.40\linewidth}
\includegraphics[width=\linewidth]{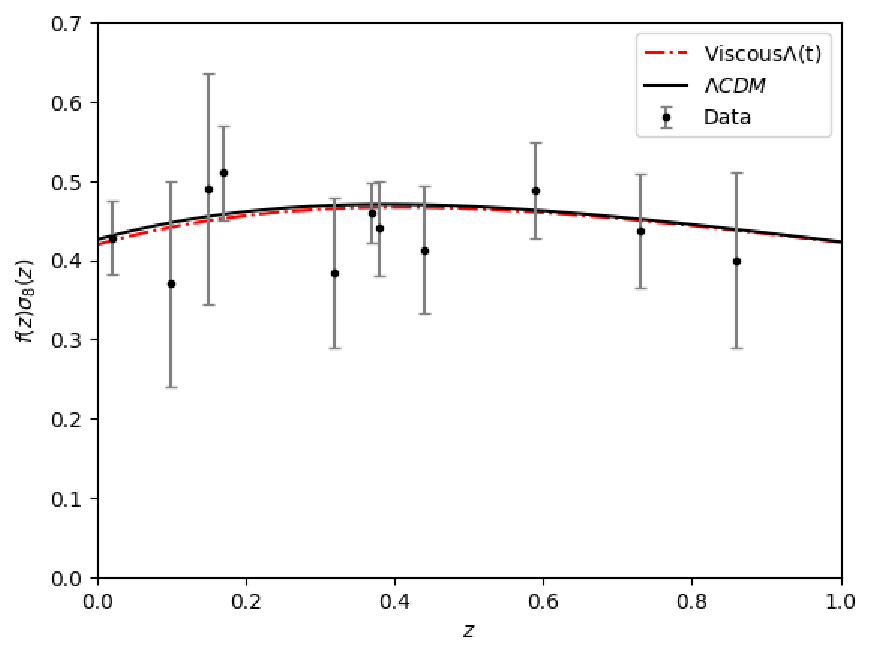}
\caption{Theoretical curves for the $f(z)\sigma_8(z)$ corresponding to $\Lambda$CDM and viscous $\Lambda(t)$ model along with some of the data points employed in our analysis. To generate this plot we have used the best-fit values of the cosmological parameters listed in Tables I and II for $``+f\sigma_8"$ data.}
\label{fig:17}
\end{minipage}
\hfill
\begin{minipage}[t]{0.40\linewidth}
\includegraphics[width=\linewidth]{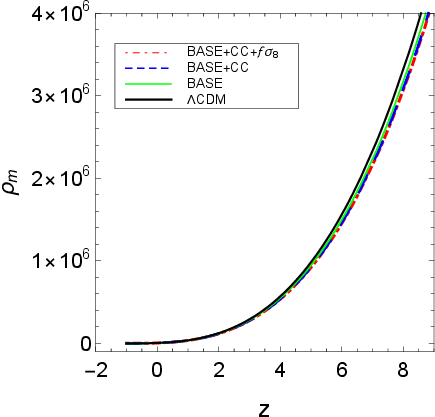}
\caption{The matter energy  density as a function of redshift for  decaying vacuum with  viscous term $\zeta=\zeta_1 H$ using the best fit values obtained from different combinations of datasets.}
\label{fig:18}
\end{minipage}
\end{figure*}

\section{Selection criterion}\label{aicbic}
\indent There are two widely used selection criterion, namely, Akaike information criteria (AIC) and Bayesian information criteria (BIC) to measure the goodness of the fitted models compared to a base model. AIC is an essentially selection criteria based on the information theory where as the BIC is based on the bayesian evidence valid for large sample size. In cosmology, AIC and BIC are used to discriminate cosmological models based on the penalization associated with the number of free parameters of the considered models. The AIC parameter is defined through the relation \cite{akaike1974}
\begin{equation}
AIC = \chi^{2}_{min} + \frac{2dN}{N-d-1},
\end{equation}
where $d$ is the free parameters in a model, $N$ the observational data points and $\chi^2_{min}$ is the minimum value of the $\chi^2$ function. AIC penalizes according to the number of free parameters of that model. To discriminate the proposed model $m_1$ with the reference model $m_2$, we calculate $\Delta AIC_{m_1m_2}=AIC_{m_1}-AIC_{m_2}$, which can be explained as ``evidence in favor" of model $m_1$ as compared to model $m_2$. In this paper, we consider $\Lambda$CDM model as a reference model ($m_2$). \\
\indent The value $0\le \Delta AIC_{m_1m_2}<2$ refers to ``strong evidence in favor" of the model $m_1$, for $2\le \Delta AIC_{m_1m_2}\le4$, there is ``average strong evidence in favor" of the model $m_1$, for $4< \Delta AIC_{m_1m_2}\le7$, there is ``little evidence in favor" of the model $m_1$, and for $\Delta AIC_{m_1m_2}>8$ there is ``no evidence in favor" of the model $m_1$.\\
\indent On the other hand, the Bayesian information criteria (BIC) can be defined as  \cite{schwarz1978}
\begin{equation}
BIC = \chi^{2}_{min} + d\ln N.
\end{equation}
\indent Similar to $\Delta AIC$, $\Delta BIC_{m_1m_2}=BIC_{m_1}-BIC_{m_2}$ gives as ``evidence against" the model $m_1$ with reference to model $m_2$. For $0 \le \Delta BIC_{m_1m_2} < 2$ gives ``not enough evidence" of the model $m_1$, for $2 \le \Delta BIC_{m_1m_2} < 6$, we have ``evidence against" the model $m_1$, and for $6 \le \Delta BIC_{m_1m_2} < 10$, there is ``strong evidence against"  the model $m_1$. Finally, if $\Delta BIC > 10$ then there is strong evidence against the model and it is probably not the best model.\\
\indent The values of $\Delta$AIC and $\Delta$BIC with respect to $\Lambda$CDM as the referring model are shown in Table III. According to our results, $\Delta AIC (\Delta BIC)$ = $1.026 (10.977)$ with respect to the $BASE$ dataset, $\Delta AIC (\Delta BIC)$ = $0.959 (10.913)$ with $+CC$ dataset, and for $+f\sigma_8$ dataset, we have $\Delta AIC (\Delta BIC)$ = $-7.492 (2.416)$. Thus, under AIC there is ``strong evidence in favor" of the viscous $\Lambda(t)$ model where as under BIC, there is ``strong evidence against" the viscous $\Lambda(t)$ model with  $BASE$ and $+CC$ dataset and ``positive evidence against" the model with  $+f\sigma_8$ dataset.
\section{Conclusion}\label{conclusion}
In this work, we have studied the analytical and observational consequences of cosmology inspired by dissipative phenomena in fluids according to Eckart theory with varying VED scenarios for spatially flat homogeneous and isotropic FLRW geometry. We have assumed the interaction of two components: viscous dark matter and vacuum energy density satisfying the conservation equation (8). we have solved the field equations by assuming the most general form of bulk viscous coefficient, viz., $\zeta=\zeta_0+\zeta_1H+\zeta_2 (\ddot{a}/aH)$. We have also explored three particular cases of bulk viscosity, namely (1) $\zeta=\zeta_0$; (2) $\zeta=\zeta_1 H$; (3) $\zeta=\zeta_0+\zeta_1 H$ to observe the effect of viscosity with varying VED. These viscous models have different theoretical motivations, but not all of them are able to constraint observationally. We have constrained only the viscous model $\zeta=\zeta_1 H$ with varying VED. The motivation of the present work is to study the dynamics and evolutions of a wide class of viscous models with time varying vacuum energy density in the light of the most recent observational data. Current observations do not rule out the possibility of varying DE. It has been observed that the dynamical $\Lambda$ could be useful to solve the coincidence problem. Although the functional form of $\Lambda(t)$ is still unknown, a quantum field theory (QFT) approach has been proposed within the context of the renormalization group (RG). Thus, we have used the varying VED of the functional form $\rho_{\Lambda}=c_0+3\nu H^2$ in all of viscous models presented in this paper. The motivation for this functional form stems from the general covariance of the effective action in QFT in curved geometry. It has been shown that the $\Lambda(t)$ provides either a particle production processes or increasing the mass of the viscous dark matter particles. In what follows, we summarize the main results of the four different viscous $\Lambda(t)$ models.\\
\indent In case of the viscous $\Lambda(t)$ models with $\zeta=\zeta_0$, $\zeta=\zeta_0+\zeta_1 H$ and $\zeta=\zeta_0+\zeta_1H+\zeta_2 (\ddot{a}/aH)$, we have found the analytical solutions of the various cosmological parameters, like $H(t)$, $a(t)$, $\rho_m(t)$, $q(t)$ and $w_{eff}(t)$. It has been observed that all these three viscous $\Lambda(t)$ models expand exponentially with cosmic time $t$. The models show the transition from decelerated phase to accelerated phase in late time. The matter energy density, $\rho_m(t)$ approaches to a finite value in late time evolution of the Universe. This happens due to the presence of bulk viscosity. The deceleration parameter $q(t)$ tends to $-1$ as $t\rightarrow \infty$. It is important to note that it is $H(z)$ that is actually the observable quantity in cosmology which can be examined with current observations. However, assuming suitable choice of model parameters, we have discussed numerically the evolutions and dynamics of these models. \\
\indent In case of viscous $\Lambda(t)$ model with $\zeta=\zeta_1 H$, we have obtained the various cosmological parameters. We have performed a joint likelihood analysis in order to put the constrain on the main parameters by using the three different combinations of observational data: BASE, $+CC$ and $+f\sigma_8$. To discriminate our model with the concordance $\Lambda$CDM model, we have also performed the statistical analysis for $\Lambda$CDM by using the same observational datasets. Our finding shows that this viscous $\Lambda(t)$ model can accommodate a late time accelerated expansion. It has been observed that we can improve significantly the performance of the model by using $BASE+CC+f\sigma_8$.\\
\indent From observational consistency points of view, we have examined the evolution of the viscous $\Lambda(t)$ model on Hubble parameter, deceleration parameter and equation of state parameter by using the best-fit values of parameters. It has been observed that the model depicts transition from an early decelerated phase to late-time accelerated phase and the transition takes place at $z_{tr}=0.664^{+0.031}_{-0.042}$ with $BASE$ data, $z_{tr}=0.665^{+0.031}_{-0.037}$ with $+CC$ data and  $z_{tr}=0.626^{+0.028}_{-0.037}$ with  $+f \sigma_8$ data. The present viscous $\Lambda(t)$ model has $q_0=-0.533^{+0.025}_{-0.020}$, $q_0=-0.535^{+0.023}_{-0.020}$ and $q_0=-0.516^{+0.022}_{-0.017}$ respectively. Thus, both $z_{tr}$ and $q_0$ values are in good agreement with that of $\Lambda$CDM model. The ages of the Universe obtained for this model with each dataset are very much compatible with the $\Lambda$CDM model. The proposed model has small negative value of EoS parameter for large redshifts and asymptotically approaches to cosmological constant for small redshifts. Thus, the viscous $\Lambda(t)$ model behaves like quintessence in early time and cosmological constant in late-time. The residual tensions of our fitting results with respect to the latest local measurement $H_0 = 73.04 \pm 1.04$ km/s/Mpc \cite{riess2022} are $3.92\sigma$, $3.85\sigma$ and $4.07\sigma$, respectively. In Ref. \cite{ywang2017}, the authors found $H_0=69.13\pm2.34$ km/s/Mpc assuming the $\Lambda$CDM. Such result almost coincides with $H_0$ that we obtained in Tables I and II for $\Lambda$CDM and viscous $\Lambda(t)$ models. We have explored the $\sigma_8$ and $S_8$   parameters using the combined datasets of $BASE+CC+f\sigma_8$. The constraints on $\sigma_8$ and $S_8$ from this combined analysis are $\sigma_8$ = $0.790^{+0.008}_{-0.010}$ and $S_8=0.822^{+0.019}_{-0.019}$, respectively which are very close to the values of $\Lambda$CDM. The tension of our fitting results in $\sigma_8$ and $S_8$ for viscous $\Lambda(t)$ model with respect to respective $\sigma_8$ and $S_8$ of $\Lambda$CDM are $0.23\sigma$ and $-0.38\sigma$, respectively. The evolution of $f\sigma_8$ as displayed in Fig.17 shows that the behaviour of $f\sigma_8$ is consistent with the observational data points. It has been noticed that the best-fit results are consistent in the vicinity of Planck data \cite{agh2020}.  \\
\indent  It has been observed that the value of $\chi^2_{red}$ is less than unity with every data sets which show that the model is in a very good fit with these observational data sets and the observed data are consistent with the considered model. The jerk parameter remains positive and less than unity in past, and eventually tends to unity in late-time. Thus, the jerk parameter deviates in early time but it attains the same value as $\Lambda$CDM in late-time.\\
\indent To discriminate the viscous $\Lambda(t)$ with the $\Lambda$CDM, we have examined the selection criterion, namely, AIC and BIC. According to the selection criteria $\Delta$AIC, we have found that the viscous $\Lambda(t)$ model is ``positively favored" over the $\Lambda$CDM model for $BASE$, $+CC$ and $+f\sigma_8$ datasets. Similarly, with respect to $\Delta$BIC our model has a ``very strong evidence against" the model for $BASE$ and $+CC$ datasets whereas when we add $+f\sigma_8$ dataset, there is ``no significant evidence against" the model. As a concluding remark we must point out that the viscous models with decaying VED may be preferred as potential models to examine the dark energy models beyond the concordance cosmological constant. The viscous effects with decaying VED can drive an accelerated expansion of the Universe. Thus, a viable cosmology can be constructed with viscous fluids and decaying VED. With new and more accurate observations, and with more detailed analyses, it would be possible to conclusively answer the compatibility of viscous model with dynamical vacuum energy.
\begin{acknowledgements}
One of the author, VK would like to thank Delhi Technological University, India for providing Research Fellowship to carry out this work.
\end{acknowledgements}

\end{document}